\documentclass[10pt,journal,a4paper,compsoc]{IEEEtran}

\usepackage{cite}

\usepackage[pdftex]{graphicx}
\graphicspath{{figures/}}
\DeclareGraphicsExtensions{.pdf,.jpeg,.png}

\usepackage[cmex10]{amsmath}
\usepackage{amsthm}
\usepackage{url}
\usepackage{textcomp}

\usepackage[OT1]{fontenc}
\usepackage{preamble}
\usepackage[french,english]{babel}

\usepackage{xcolor}
\usepackage{multicol}
\usepackage{kbordermatrix}
\renewcommand{\kbldelim}{(}
\renewcommand{\kbrdelim}{)}

\newtheorem{proposition}{Proposition}

\newtheorem{lemma}{Lemma}
\newtheorem{theorem}{Theorem}

\newcommand{\cl}[1]{\mathcal{#1}}
\newcommand{\clM}{\cl{M}}

\makeatletter
\let\l@ENGLISH\l@english
\makeatother

\begin{document}
\title{Structured regularization for conditional Gaussian graphical models}

\author{Julien~Chiquet, Tristan~Mary-Huard and~St\'ephane~Robin,
  \IEEEcompsocitemizethanks{
    \IEEEcompsocthanksitem     J.     Chiquet,
    T.   Mary-Huard   and   S.    Robin   are   with   the   UMR   518
    AgroParisTech/INRA,    rue    Claude    Bernard,   70005    Paris,
    FRANCE. \protect\\
    E-mail: {chiquet,robin,maryhuar@agroparistech.fr}
    \IEEEcompsocthanksitem   T.   Mary-Huard  is   with   the  UMR   de
    G\'en\'etique  V\'eg\'etale du  Moulon, INRA/Univ.  Paris Sud/CNRS,
    Ferme du Moulon, 91190 Gif-sur-Yvette
  \IEEEcompsocthanksitem J. Chiquet is with the LaMME - UMR 8071 CNRS/UEVE,
  Boulevard de France, 91000, \'Evry, FRANCE.}
    \thanks{}}

\markboth{
}%
{Chiquet \MakeLowercase{\textit{et al.}}: Structured Regularization for cGGM}


\IEEEcompsoctitleabstractindextext{%
  \begin{abstract}
    Conditional  Gaussian   graphical  models  (cGGM)   are  a  recent
    reparametrization  of  the  multivariate linear  regression  model
    which explicitly exhibits $i)$ the partial covariances between the
    predictors and  the responses,  and $ii)$ the  partial covariances
    between  the responses  themselves. Such  models  are particularly
    suitable for  interpretability since partial  covariances describe
    strong  relationships between  variables.  In  this  framework, we
    propose a  regularization scheme to enhance  the learning strategy
    of  the model  by  driving  the selection  of  the relevant  input
    features  by  prior  structural  information.  It  comes  with  an
    efficient alternating  optimization procedure which  is guaranteed
    to converge to the global  minimum.  On top of showing competitive
    performance  on   artificial  and  real  datasets,  our  method
    demonstrates   capabilities  for   fine   interpretation  of   its
    parameters,  as illustrated  on three  high-dimensional  datasets
    from spectroscopy, genetics, and genomics.
\end{abstract}

}

\maketitle

\IEEEdisplaynotcompsoctitleabstractindextext
\IEEEpeerreviewmaketitle

\section{Introduction}

Multivariate regression,  i.e.  regression with  multiple response
variables, is increasingly used to model high dimensional problems. By
considering multiple  responses, we wish to  strengthen the estimation
and/or selection  of  the relevant  input  features,  by  taking
advantage  of the  dependency  pattern between  the  outputs. This  is
particularly appealing when the data is scarce, or even in the '$n<p$'
high dimensional setup, in which framework this work enters.  Compared
to  its  univariate counterpart,  the  general  linear model  aims  to
predict several -- say $q$ --  responses from a set of $p$ predictors,
relying on a training data set $\set{(\bx_i,\by_i)}_{i=1,\dots,n}$:
\begin{equation}
  \label{eq:reg_model_row}
  \by_i      =     \bB^T      \bx_i     +      \bvarepsilon_i,     \quad
  \bvarepsilon_i\sim\mathcal{N}(\bzr,\bR), \quad \forall i=1,\dots,n.
\end{equation}
The  $p\times  q$  matrix  of  regression coefficients  $\bB$  and  the
$q\times   q$  covariance   matrix   $\bR$  of   the  Gaussian   noise
$\bvarepsilon_i$ are unknown.
Model \eqref{eq:reg_model_row} has been studied by \cite{MV_mardia} in
the  low dimensional case  where both  ordinary and  generalized least
squares estimators of $\bB$ coincide and do not depend on $\bR$. These
approaches  boil down  to  performing $q$  independent regressions,  each
column $\bB_j$  describing the weights associating  the $p$ predictors
to the $j$th  response.  In the $n<p$ setup  however, these estimators
are not defined.

Mimicking the  univariate-output case, multivariate  penalized methods
aim to regularize  the problem by biasing  the regression coefficients
toward a given feasible set.  Sparsity within the set of predictors is
usually the most wanted feature in the high-dimensional setting, which
can  be  met  in  the  multivariate  framework  by  a  straightforward
application  of  the  most  popular  penalty-based  methods  from  the
univariate   world  involving   $\ell_1$-regularization.   Still,   by
encouraging sparsity  or any  other priors  by not  distinguishing the
regression   parameters  across   the   outputs,   we  roughly   treat
multivariate regression  as independent multiple  regression problems.
It is obviously of the highest  interest to treat the problem jointly,
so that the structure of the  output drives the way the parameters are
penalized, and eventually, selected.  Several recent papers tackle the
multivariate  regression  problem  with  penalty-based  approaches  by
including  the  dependency pattern  of  the  outputs in  the  learning
process.  When  this pattern  is known \emph{a  priori}, we  enter the
``multitask'' framework where many  authors have suggested variants of
the  $\ell_1$-penalty shaped  accordingly.  A  natural approach  is to
encourage a similar  behavior of the regression  parameters across the
outputs -- or  tasks -- using group-norms zeroing a  full row of $\bB$
at   once,    across   the    corresponding   outputs    (see,   e.g.,
\cite{2011_SC_Chiquet,2011_AOS_Obozinski}).  Refinements exist to cope
with  complex dependency  structures, using  graph or  tree structures
(\cite{2009_PLOS_kim,2010_ICML_kim}),  yet  the  pattern  between  the
responses remains fixed.  When it is unknown, the general linear model
\eqref{eq:reg_model_row} is a natural tool to account for dependencies
between        the       responses.         In       this        vein,
\cite{2010_JCGS_rothman,2011_AoAS_yin} suggest penalizing the negative
log-likelihood  of  \eqref{eq:reg_model_row}  by  two  $\ell_1$  norms
respectively inducing  sparsity on  the regression  coefficients $\bB$
and on the inverse  covariance $\bR^{-1}$.  Theoretical guarantees are
proposed in \cite{2012_JMA_lee}, with a two-stage procedure
involving a plug-in  estimator of the inverse  covariance, which leads
to a  computationally less demanding optimization  procedure.  However
their  criterion is  hard  to  minimize as  it  is  only bi-convex  in
$(\bB,\bR^{-1})$, and  no theoretical guarantee is  provided regarding
the convergence of the proposed optimization strategy.

In       \cite{2012_JMLR_sohn,2014_ITIEEE_yuan},       an      elegant
re-parametrization of \eqref{eq:reg_model_row} is depicted, leading to
a formulation that  is jointly convex and enjoys  features of Gaussian
Graphical   Models  (GGM).    This   has  been   referred   to  as   a
\textit{'conditional    Gaussian   Graphical    Model'}    (cGGM)   or
\textit{'partial Gaussian Graphical  Model'} in the recent literature.
A remarkable feature of this  formulation is to explicitly exhibit the
\emph{direct} relationships that  exist between the predictors $\bx_i$
and the  responses $\by_i$ through their  partial covariances (denoted
$\bOmega_{\bx\by}$   hereafter).   Dealing   with  the   direct  links
$\bOmega_{\bx\by}$ is much  more convenient for interpretability than is
dealing with the regression parameters $\bB$, the latter entailing both
direct  and   indirect  influences,   possibly  due  to   some  strong
correlations between the  responses.  Regarding the learning strategy,
\cite{2012_JMLR_sohn} propose to regularize the cGGM log-likelihood by
two  $\ell_1$-norms  respectively acting  on  the partial  covariances
between the features and the responses $\bOmega_{\bx\by}$ on the first
hand, and  on the partial covariance between  the responses themselves
via $\bR^{-1}$  on the second hand.

In this paper,  we build on this approach yet  in a slightly different
setting.  First we  typically assume  a reasonable  number of  outputs
compared to the number of predictors  and sample size (while we insist
on  the fact  that  the number  of predictors  may  exceed the  sample
size). As a consequence our regularizer induces sparsity on the direct
relationships   like  in   \cite{2012_JMLR_sohn},   yet  no   sparsity
assumption is  made for  the inverse  covariance. Second,  we consider
applications   where   structural   information  about   sparsity   is
available. Here  the structural  information will  be embedded  in the
regularization scheme  via an additional regularization  term using an
application-specific metrics, in the same manner as in the 'structured'
versions               of                the               Elastic-net
\cite{2010_AOAS_slawski,2011_EJS_Hebiri,2010_AISTATS_Lobert},       or
quadratic    penalty    function    using    the    Laplacian graph
\cite{2007_BF_Rapaport,2010_AOAS_Li} proposed in the univariate-output
case.  We  show that  the resulting penalized  likelihood optimization
problem can be solved  efficiently and provide algorithmic convergence
guarantees  for  the  two-step  procedure  presented  here.  Penalized
criteria for  the choice  of the  regularization parameters  are also
provided.  We  also  investigate   the  importance  of  embedding  for
structural prior information in  the various contexts of spectroscopy,
genomic  selection and  regulatory motif  discovery, illustrating  how
accounting  for  application-specific  improves both  performance  and
interpretability.  The procedure is available as an \texttt{R}-package
called \textbf{spring}, available on the \texttt{R-forge}.

The outline of the paper is as follows.  In Section \ref{Sec:model} we
provide background  on cGGM and  present our regularization  scheme. In
Section  \ref{Sec:learning},  we  develop  an  efficient  optimization
strategy in order to  minimize the  associated  criterion.   A paragraph  also
addresses  the  model  selection  issue.   Section  \ref{Sec:simu}  is
dedicated   to   illustrative    simulation   studies.    In   Section
\ref{Sec:applications}, we  investigate three multivariate  data sets:
first, we  consider an  example in spectrometry  with the  analysis of
cookie  dough  samples;  second,  the  relationships  between  genetic
markers and a series of  phenotypes of the plant \emph{Brassica napus}
is addressed;  and third, we  investigate the discovery  of regulatory
motifs of yeast from time-course microarray experiments.   In these
applications, some specific underlying structuring priors arise, the
integration of which within  our model  is detailed  as it is  one of  the main
contributions of our work.


\section{Model setup}
\label{Sec:model}
\subsection{Background on cGGM}
\label{Sec:stat_model}

The   statistical  framework   of   cGGM  arises   from  a   different
parametrization  of  \eqref{eq:reg_model_row}   which  fills  the  gap
between  multivariate   regression  and   GGM.   It  extends   to  the
multivariate case the links existing between the linear model, partial
correlations    and    GGM,    as    depicted    for    instance    by
\cite{2006_AS_Meinshausen} then \cite{2009_JASA_Peng}.  To the best of
our knowledge,  the connection  in the case  of multiple  reponses was
first underlined  by \cite{2012_JMLR_sohn},  which we recall  here.  It
amounts to investigating the  joint probability distribution of $(\bx_i,
\by_i)$  in   the  Gaussian   case,  with  the   following  block-wise
decomposition  of  the covariance  matrix  $\bSigma$  and its  inverse
$\bOmega=\bSigma^{-1}$:
  \begin{equation}
    \label{eq:joint_model}
    \bSigma = \begin{pmatrix}
      \bSigma_{\bx\bx} & \bSigma_{\bx\by} \\
      \bSigma_{\by\bx} & \bSigma_{\by\by} \\
   \end{pmatrix},\quad
   \bOmega = \begin{pmatrix}
     \bOmega_{\bx\bx} & \bOmega_{\bx\by} \\
     \bOmega_{\by\bx} & \bOmega_{\by\by} \\
   \end{pmatrix}.
 \end{equation}
 Back  to   the  distribution  of  $\by_i$   conditional  on  $\bx_i$,
 multivariate Gaussian  analysis shows that, for  centered $\bx_i$ and
 $\by_i$,
\begin{equation}
  \label{eq:cond_model}
  \by_i | \bx_i \sim
  \mathcal{N}\left(-\bOmega^{-1}_{\by\by}\bOmega_{\by\bx}
    \bx_i, \bOmega_{\by\by}^{-1} \right).
\end{equation}
This      model,      associated     with     the      full      sample
$\{(\bx_i,\by_i)\}_{i=1,\dots,n}$, can be written  in a matrix form by
stacking in  rows first the observations of  the responses,
and then of the  predictors, in  two data matrices $\bY$
and $\bX$ with respective sizes $n\times q$ and $n\times p$, such that
\begin{multline}
  \label{eq:cond_model_mat}
  \bY = -\bX \bOmega_{\bx\by} \bOmega_{\by\by}^{-1} + \bvarepsilon, \\
  \vec(\bvarepsilon) \sim
  \mathcal{N}\left(\bzr_{nq}, \bI_{n} \otimes \bOmega_{\by\by}^{-1} \right),
\end{multline}
where, for a  $n \times p$ matrix ${\bf A}$,  denoting ${\bf A}_j$ its
$j$th  column, the  $\vec$ operator  is  defined as  $\vec({\bf A})  =
(\bA_1^T     \dots     \bA_p^T)^T$.      The     log-likelihood     of
\eqref{eq:cond_model_mat}  --   which  is  a   conditional  likelihood
regarding the joint model \eqref{eq:joint_model} -- is written
\begin{multline}
  \label{eq:log_lik}
  \log L(\bOmega_{\bx\by},\bOmega_{\by\by}) =
  \frac{n}{2}\log\left|\bOmega_{\by\by}\right| + \\ \frac{n}{2}
  \trace\left(  \left(\bY +  \bX \bOmega_{\bx\by}\bOmega_{\by\by}^{-1}
    \right)          \bOmega_{\by\by}\left(\bY          +          \bX
      \bOmega_{\bx\by}\bOmega_{\by\by}^{-1} \right)^T\right) +
  \mathrm{cst.}
\end{multline}
\begin{figure*}[htbp!]
  \centering
  \begin{tabular}{c@{\hspace{1cm}}c}
    \begin{tabular}{cccc}
      & $\bR_{\text{low}}$ & $\bR_{\text{med}}$ & $\bR_{\text{high}}$ \\
      & \includegraphics[scale=.1]{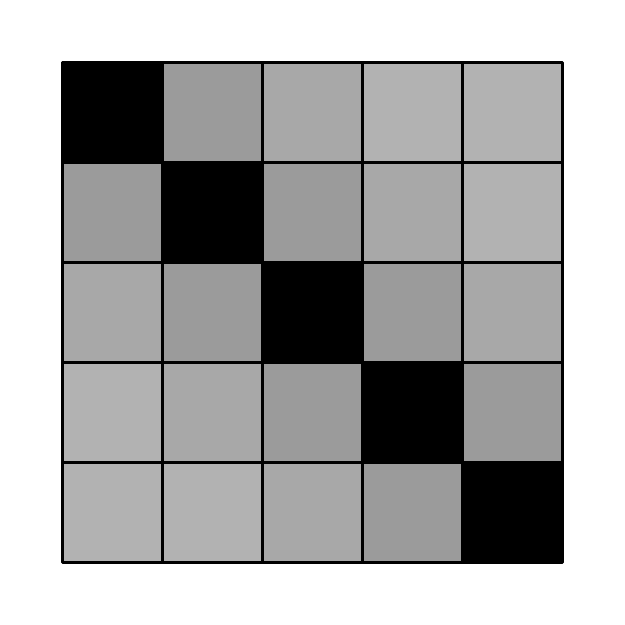}
      & \includegraphics[scale=.1]{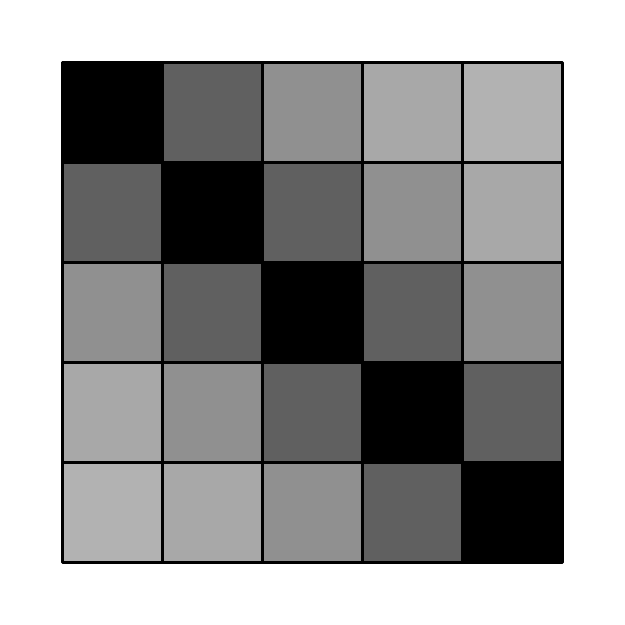}
      & \includegraphics[scale=.1]{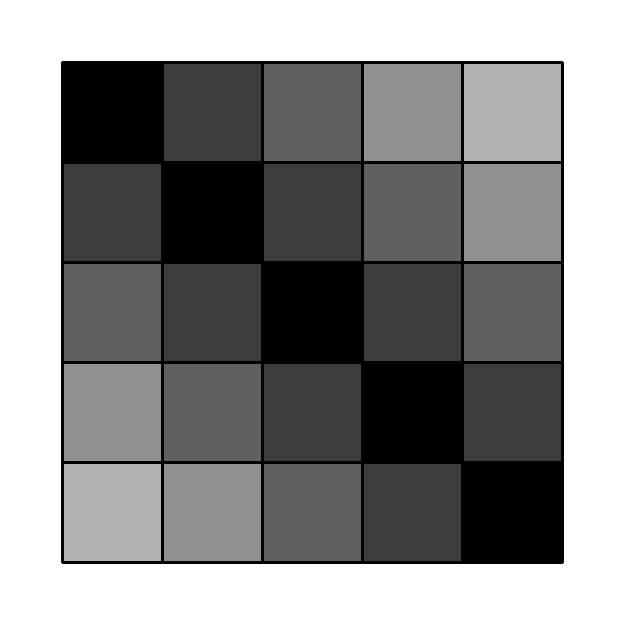} \\
      $\bOmega_{\bx\by}$ & $\bB_{\text{low}}$ & $\bB_{\text{med}}$ & $\bB_{\text{high}}$ \\
      \includegraphics[scale=.12]{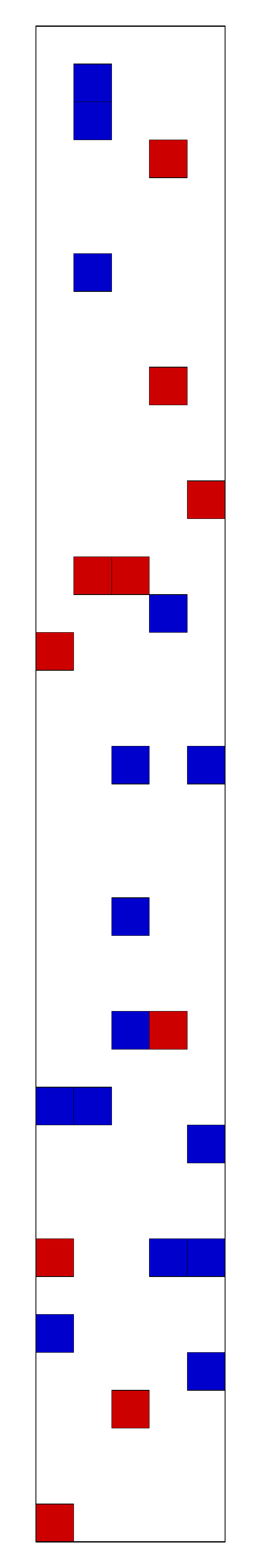}
      & \includegraphics[scale=.12]{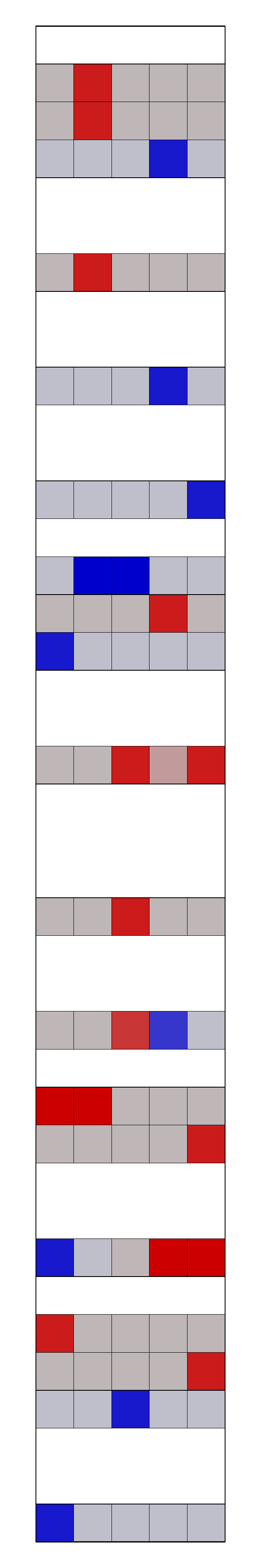}
      & \includegraphics[scale=.12]{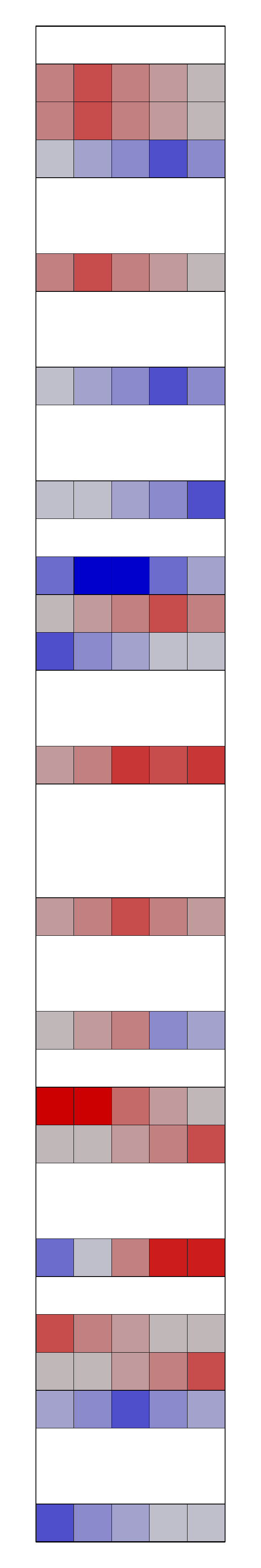}
      & \includegraphics[scale=.12]{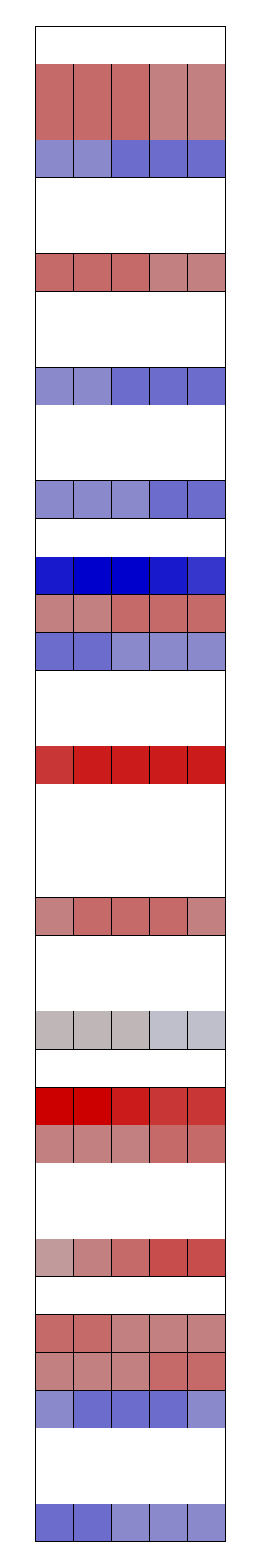}\\
    \end{tabular}
    &
    \begin{tabular}{cccc}
      & $\bR_{\text{low}}$ & $\bR_{\text{med}}$ & $\bR_{\text{high}}$ \\
      & \includegraphics[scale=.1]{figures/R_low_cov}
      & \includegraphics[scale=.1]{figures/R_med_cov}
      & \includegraphics[scale=.1]{figures/R_high_cov} \\
      $\bOmega_{\bx\by}$ & $\bB_{\text{low}}$ & $\bB_{\text{med}}$ & $\bB_{\text{high}}$ \\
      \includegraphics[scale=.12]{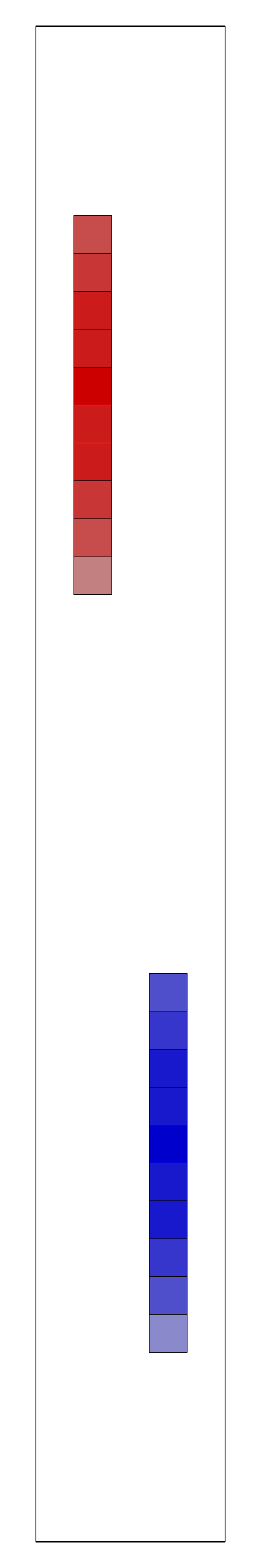}
      & \includegraphics[scale=.12]{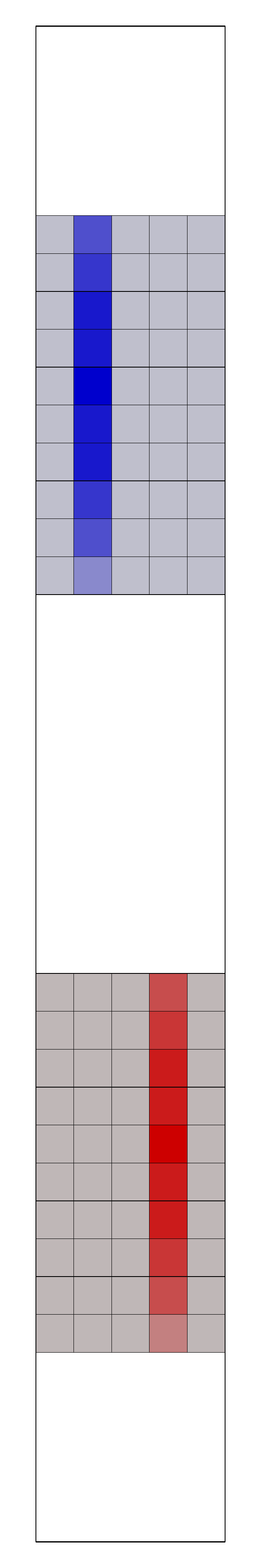}
      & \includegraphics[scale=.12]{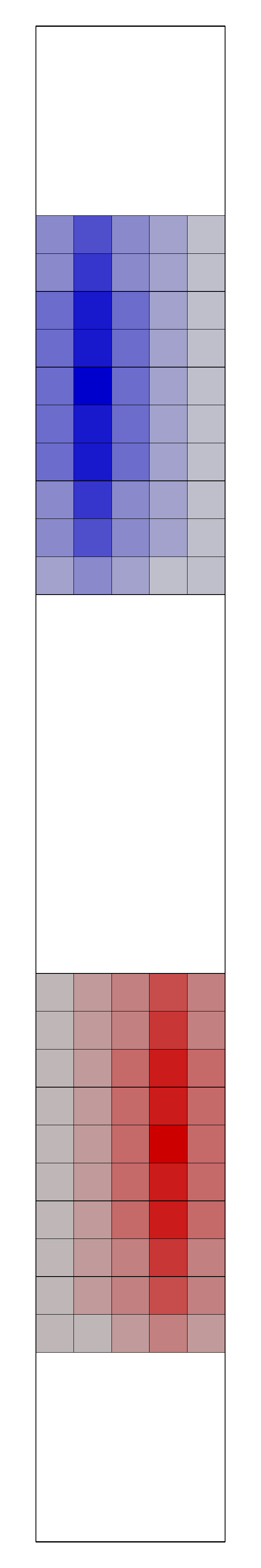}
      & \includegraphics[scale=.12]{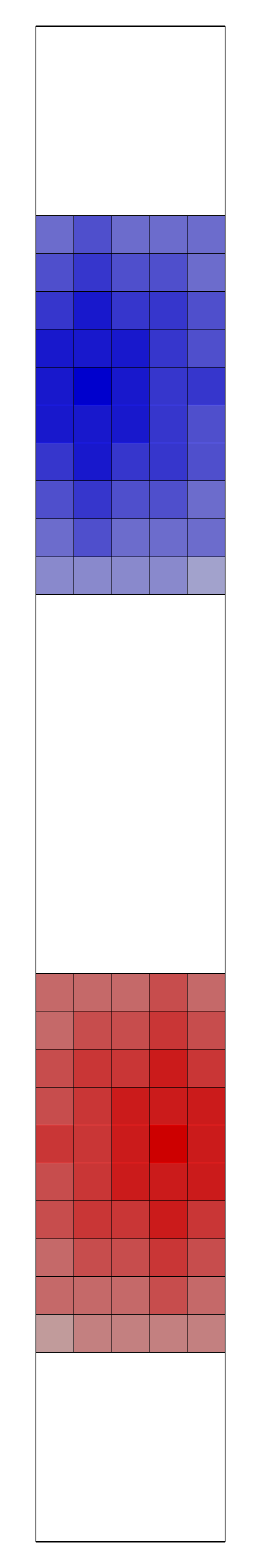}\\
    \end{tabular} \\
    (a) & (b) \\
  \end{tabular}
  \caption{Toy examples to  illustrate the relationships between $\bB,
    \bOmega_{\bx\by}$ and  $\bR$ in the cGGM (better  seen in color):
    on panel $a)$, a situation  with no particular structure among the
    predictors; on  panel $b)$, a strong  neighborhood structure.  For
    each panel,  we represent the  effect of stronger  correlations in
    $\bR$ on masking the direct links in $\bB$.}
  \label{fig:model_insight}
\end{figure*}
Introducing  the  empirical  matrices  of covariance  $\bS_{\by\by}  =
n^{-1}\sum_{i=1}^n\by_i        \by_i^T$,        $\bS_{\bx\bx}        =
n^{-1}\sum_{i=1}^n\bx_i     \bx_i^T$,      and     $\bS_{\by\bx}     =
n^{-1}\sum_{i=1}^n\by_i \bx_i^T$, one has
\begin{multline}
  \label{eq:log_lik2}
  - \frac{2}{n}\log L(\bOmega_{\bx\by},\bOmega_{\by\by}) =
  - \log\left|\bOmega_{\by\by}\right| +
  \trace\left(\bS_{\by\by} \bOmega_{\by\by} \right) \\
  +     2\trace\left(    \bS_{\bx\by}\bOmega_{\by\bx}     \right)    +
  \trace(\bOmega_{\by\bx}\bS_{\bx\bx}                  \bOmega_{\bx\by}
  \bOmega_{\by\by}^{-1}) + \mathrm{cst.}
\end{multline}
We  notice   by  comparing  the  cGGM   \eqref{eq:cond_model}  to  the
multivariate    regression    model   \eqref{eq:reg_model_row}    that
$\bOmega_{\by\by}^{-1}  = \bR  \text{  and }  \bB =  -\bOmega_{\bx\by}
\bOmega_{\by\by}^{-1}$.          Although         equivalent        to
\eqref{eq:reg_model_row},      the     alternative     parametrization
\eqref{eq:cond_model}  shows two  important  differences with  several
implications.   First, in  light  of convex  optimization theory,  the
negative log-likelihood \eqref{eq:log_lik} can  be shown to be jointly
convex in $(\bOmega_{\bx\by},\bOmega_{\by\by})$ (a formal proof can be
found  in  \cite{2014_ITIEEE_yuan}).   Minimization problems  involving
\eqref{eq:log_lik} will  thus be amenable to a  global solution, which
facilitates both optimization and  theoretical analysis.  As such, the
conditional negative log-likelihood \eqref{eq:log_lik} will serve as a
building  block for our  learning criterion.   Second, it  unveils new
interpretations  for  the   relationships  between  input  and  output
variables,  as discussed in  \cite{2012_JMLR_sohn}: $\bOmega_{\bx\by}$
describes  the  \emph{direct}  relationships  between  predictors  and
responses, the support of which  we are looking for to select relevant
interactions.  On the other  hand, $\bB$ entails \emph{both direct and
  indirect}  influences,  possibly  due  to some  strong  correlations
between the  responses, described by  the covariance matrix  $\bR$ (or
equivalently its  inverse $\bOmega_{\by\by}$).  To  provide additional
insights  on  cGGM,  Figure  \ref{fig:model_insight}  illustrates  the
relationships between  $\bB,\bOmega_{\bx\by}$ and $\bR$  in two simple
scenarios  where  $p=40$  and  $q=5$.   Scenarios $a)$  and  $b)$  are
discriminated  by  the  presence  of  a  strong  structure  among  the
predictors. Still, the  important point to grasp at  this stage of the
paper  is  how strong  correlations  between  outcomes can  completely
``mask'' the direct links in the regression coefficients: the stronger
the correlation  in $\bR$, the less  it is possible  to distinguish in
$\bB$ the non-zero coefficients of $\bOmega_{\bx\by}$.


\subsection{Structured regularization with underlying sparsity}
\label{Sec:crit_learn}

Our regularization scheme starts  by considering some structural prior
information about the  relationships between the coefficients. We are
typically thinking of a situation where similar inputs are expected to
have  similar direct relationships  with the  outputs. The right  panel of
Figure  \ref{fig:model_insight}   represents  an illustration of  such  a
situation,  where  there  exists  an  extreme  neighborhood  structure
between the  predictors.  This depicts  a pattern that acts  along the
rows of $\bB$ or  $\bOmega_{\bx\by}$ as substantiated by the following
Bayesian point of view.

\subsubsection*{Bayesian interpretation}  Suppose that the similarities  can be
encoded  into  a  matrix  $\bL$.   Our  aim is  to  account  for  this
information  when learning the  coefficients.  The  Bayesian framework
provides  a  convenient  setup  for  defining  the  way  the  structural
information  should  be accounted  for.   In  the  single output  case
(see, e.g. \cite{MaR07}), a conjugate prior for $\bbeta$ would be
$\mathcal{N}(\bzr, \bL^{-1})$.   Combined with the  covariance between
the outputs, this gives
\begin{equation*}
  \vec(\bB) \sim \mathcal{N}(\bzr,\bR \otimes \mathbf{L}^{-1}),
\end{equation*}
where $\otimes$ is the Kronecker  product. By properties of the $\vec$
operator, 
this can be stated straightforwardly as
\begin{equation*}
  \vec(\bOmega_{\bx\by}) \sim \mathcal{N}(\bzr,\bR^{-1} \otimes \mathbf{L}^{-1})
\end{equation*}
for the direct links. Choosing such a prior results in
\begin{equation*}
  \label{eq:log_lik_prior}
  \log \prob(\bOmega_{\bx\by}|\mathbf{L},\bR) =
  \frac{1}{2}\trace\left(\bOmega_{\bx\by}^T             \mathbf{L}
    \bOmega_{\bx\by} \bR \right)
  + \mathrm{cst.}
\end{equation*}

\subsubsection*{Criterion} Through  this argument, we  propose a criterion
with two  terms to regularize the  conditional negative log-likelihood
\eqref{eq:log_lik}:  first,  a  smooth   trace  term  relying  on  the
available structural information  $\mathbf{L}$; and second, an $\ell_1$
norm that  encourages sparsity  among the direct  links. We  write the
criterion  as  a  function of  $(\bOmega_{\bx\by},  \bOmega_{\by\by})$
rather than $(\bOmega_{\bx\by}, \bR)$,  although equivalent in terms of
estimation,  since the  former  leads to  a  convex formulation.   The
optimization problem turns to the minimization of
\begin{multline}
  \label{eq:objective}
  J(\bOmega_{\bx\by},\bOmega_{\by\by})        = -\frac{1}{n}\log
  L(\bOmega_{\bx\by},\bOmega_{\by\by}) \\ + \frac{\lambda_2}{2}
  \trace\left(\bOmega_{\by\bx}\mathbf{L}\bOmega_{\bx\by}\bOmega_{\by\by}^{-1}\right)
  + \lambda_1 \|\bOmega_{\bx\by}\|_1.
\end{multline}

\subsection{Connection to other sparse methods}

To get more insight into our model and to facilitate connections with
existing   approaches,  we   shall   write   the  objective   function
\eqref{eq:objective}  as  a   penalized  \emph{univariate}  regression
problem.       This      amounts      to      ``vectorizing''      model
\eqref{eq:cond_model_mat} with respect to $\bOmega_{\bx\by}$, i.e., to
writing  the objective  as  a  function of  $(\bomega,\bOmega_{\by\by})$
where  $\bomega=\vec(\bOmega_{\bx\by})$.    This  is  stated   in  the
following  proposition,  which  can be  derived  from  straightforward
matrix algebra,  as proved in Appendix  Section \ref{sec:vec_obj}. The
main interest of this proposition  will become clear when deriving the
optimization  procedure  that aims at   minimizing  \eqref{eq:objective},  as  the
optimization  problem  when $\bOmega_{\by\by}$  is  fixed  turns to  a
generalized Elastic-Net  problem.  Note  that we use  $\bA^{1/2}$ to
denote  the square  root  of  a matrix,  obtained  for  instance by  a
Cholesky  factorization  in the case  of a  symmetric  positive  definite
matrix.
\begin{proposition}     \label{prop:vec_obj}     Let    $\bomega     =
  \vec(\bOmega_{\bx\by})$.     An    equivalent    vector   form    of
  \eqref{eq:objective} is
  \begin{multline*}
    \label{eq:objective_vec}
    J(\bomega,      \bOmega_{\by\by})     =      -\frac{1}{2}     \log
    |\bOmega_{\by\by}|   +    \frac{1}{2   n}   \left\|\tilde{\by}   +
      \tilde{\bX} \bomega
    \right\|_2^2 \\ + \frac{\lambda_2}{2} \bomega^T \left(\bOmega_{\by\by}^{-1} \otimes \bL
    \right) \bomega + \lambda_1 \left\| \bomega \right\|_1,
  \end{multline*}
  where the $n\times q$ dimensional vector $\tilde{\by}$ and the $n q
  \times p q$ dimensional matrix $\tilde{\bX}$ depends on
  $\bOmega_{\by\by}$ such that
  \begin{equation*}
    \tilde{\by} = \vec(\bY  \bOmega_{\by\by}^{1/2}), \quad \tilde{\bX} =
    \left(\bOmega_{\by\by}^{-1/2} \otimes \bX\right).
  \end{equation*}
\end{proposition}
When there is only one response ($q=1$) the variance-covariance matrix
$\bR$ and its inverse $\bOmega_{\by\by}$  turns to a scalar denoted by
$\sigma^2$,  the  matrix  $\bB$  to a  $p$-vector  $\bbeta$,  and  the
objective \eqref{eq:objective} can be rewritten
%
\begin{equation}
  \label{eq:criteria2_univariate}
  \log(\sigma)             +
  \frac{1}{2n\sigma^2}\left\|\by-\bX\bbeta       \right\|_2^2      +
  \frac{\lambda_2}{2\sigma^2}             \bbeta^T\bL\bbeta            +
  \frac{\lambda_1}{\sigma^2} \|\bbeta \|_1.
\end{equation}

We recognize in  \eqref{eq:criteria2_univariate} the log-likelihood of
a  ``usual''  linear  model  with  Gaussian  noise,  unknown  variance
$\sigma^2$  and  regression parameters  $\bbeta$,  plus  a mixture  of
$\ell_1$  and $\ell_2$  norms.  Assuming that  $\sigma$  is known,  we
recognize the  general structured Elastic-net regression.  If $\sigma$
is unknown,  letting $\lambda_2=0$  in \eqref{eq:criteria2_univariate}
leads to  the $\ell_1$-penalized mixture regression  model proposed by
\cite{2010_Test_Stadler},  where  both  $\bbeta$  and  $\sigma^2$  are
inferred.   In \cite{2010_Test_Stadler},  the  authors  insist on  the
importance of penalizing the vector  of coefficients by an amount that
is inversely proportional to $\sigma$: in the closely related Bayesian
Lasso  framework of  \cite{2008_JASA_Park}, the  prior on  $\bbeta$ is
defined conditionally on $\sigma$ to guarantee an unimodal posterior.


\section{Learning}
\label{Sec:learning}
\subsection{Optimization}
\label{Sec:optim}
In the classical framework of parametrization \eqref{eq:reg_model_row}, alternate strategies where optimization is  successively performed over $\bOmega_{\bx\by}$ and $\bOmega_{\by\by}$ have been proposed \cite{2010_JCGS_rothman,2011_AoAS_yin}. They come with no guarantee of convergence to the global optimum since the objective is  only  bi-convex. In the cGGM framework of \cite{2012_JMLR_sohn,2014_ITIEEE_yuan}, the optimized criterion   is  jointly convex yet no  convergence result is provided regarding the optimization procedure proposed by the authors. Here we also consider the alternate strategy for which theoretical guarantees are provided by the following theorem:
\begin{theorem}\label{thrm:optim}
Let $n\geq q$. Criterion \eqref{eq:objective} is jointly convex in  $(\bOmega_{\bx\by},\bOmega_{\by\by})$. Moreover,  the alternate optimization
\begin{subequations}
  \begin{equation}
    \label{eq:optim_cov}
    \hat{\bOmega}_{\by\by}^{(k+1)} = \argmin_{\bOmega_{\by\by}\succ 0}
    J_{\lambda_1\lambda_2}(\hat{\bOmega}_{\bx\by}^{(k)}, \bOmega_{\by\by}),
  \end{equation}
  \begin{equation}
    \label{eq:optim_par}
    \hat{\bOmega}_{\bx\by}^{(k+1)}     =    \argmin_{\bOmega_{\bx\by}}
    J_{\lambda_1\lambda_2}(\bOmega_{\bx\by},\hat{\bOmega}_{\by\by}^{(k+1)}).
  \end{equation}
\end{subequations}
leads to the optimal solution.
\end{theorem}
The proof is given in Appendix \ref{sec:proof_theorem}, and relies on the fact that efficient procedures exist to solve the two convex sub-problems \eqref{eq:optim_cov} and \eqref{eq:optim_par}.\\

Because  our  procedure  relies  on alternating  optimization,  it  is
difficult to give either a global  rate of convergence or a complexity
bound.   Nevertheless, the  complexity of  each iteration  is easy  to
derive,  since  it  amounts  to  two well-known  problems:  the  main
computational  cost in  \eqref{eq:optim_cov} is  due to  the SVD  of a
$q\times  q$  matrix,   which  costs  $\mathcal{O}(q^3)$.   Concerning
\eqref{eq:optim_par}, it  amounts to the resolution  of an Elastic-Net
problem with $p\times  q$ variables and $n \times q$  samples.  If the
final number of nonzero entries  in $\hat{\bOmega}_{\bx\by}$ is $k$, a
good  implementation  with  Cholesky  update/downdate  is  roughly  in
$\mathcal{O}(n p q^2 k)$ (see, e.g. \cite{2012_FOT_bach}). Since we
typically assumed that $p\geq n\geq q$, the global cost of a single
iteration of the alternating scheme  is thus $\mathcal{O}(n p q^2 k)$,
and  we theoretically  can  treat  problems with  large  $p$ when  $k$
remains moderate.

Finally, we typically want to compute  a series of solutions along the
regularization path of Problem \eqref{eq:objective}, i.e.  for various
values of  $(\lambda_1,\lambda_2)$.  To this  end, we simply  choose a
grid     of     penalties     $\Lambda_1    \times     \Lambda_2     =
\set{\lambda_1^{\text{min}},  \dots,   \lambda_1^{\text{max}}}  \times
\set{\lambda_2^{\text{min}},  \dots,   \lambda_2^{\text{max}}}$.   The
process is easily  distributed on different computer  cores, each core
corresponding to a value picked in  $\Lambda_2$.  Then on each core --
i.e.  for a fix $\lambda_2\in\Lambda_2$ --  we cover all the values of
$\lambda_1\in\Lambda_1$ relying on the  warm start strategy frequently
used  to  go through  the  regularization  path of  $\ell_1$-penalized
problems.  

\subsection{Model selection and parameter tuning}
\label{Sec:selection}

Model selection amounts here to choosing a couple of tuning parameters
$(\lambda_1,\lambda_2)$ among the grid $\Lambda_1 \times
\Lambda_2
$, where  $\lambda_1$ tunes the  sparsity while $\lambda_2$  tunes the
amount  of structural regularization.   We aim to pick either the
model with  minimum prediction error, or  the one closest  to the true
model,   assuming  Equation~\eqref{eq:cond_model}  holds.    These  two
perspectives  generally do  not lead  to the  same model  choices: when
looking  for  the  model  minimizing the  prediction  error,  $K$-fold
cross-validation  is   the  recommended  option  (~\cite{Hesterberg08})
despite  its   additional  computational  cost.    Letting  $\kappa  :
\set{1,\dots,n}\to\set{1,\dots,K}$ the  function indexing the  fold to
which    observation   $i$   is    allocated,   the    CV-choices   for
$(\lambda_1^\text{cv},\lambda_2^\text{cv})$ are the ones that minimize
\begin{equation*}
  \label{eq:cv_pred}
  \frac{1}{n} \sum_{i=1}^n
  \left\| \bx_{i}^T \hat{\bB}_{-\kappa(i)}^{\lambda_1,\lambda_2}- \by_{i} \right\|_2^2,
\end{equation*}
where   $\hat{\bB}_{-\kappa(i)}^{\lambda_1,\lambda_2}$  minimizes  the
objective  \eqref{eq:objective},   once  data  $\kappa(i)$   has  been
removed. In  place of  the prediction error,  other quantities  can be
cross-validated  in  our  context,  e.g. the  negative  log-likelihood
\eqref{eq:log_lik}. In  the remainder of  the paper, however, we  only consider
cross-validating the prediction error, though.

As  an alternative  to cross-validation,  penalyzed criteria  provide a
fast way to  perform model selection and are  sometimes more suited to
the  selection of  the  true underlying  model.   Although relying  on
asymptotic derivations,  they often offer  good practical performance
when the  sample size  $n$ is  not too small  compared to  the problem
dimension.   For  penalized  methods,   a  general  form  for  various
information criteria is expressed as  a function of the likelihood $L$
(defined  by \eqref{eq:log_lik}  here)  and the  effective degrees  of
freedom:
\begin{equation*}
  \label{eq:pen_crit}
  - 2\log L(\hat{\bOmega}^{\lambda_1,\lambda_2}_{\bx\by},
  \hat{\bOmega}^{\lambda_1,\lambda_2}_{\by\by})   +   \mathrm{pen}
  \cdot \mathrm{df}_{\lambda_1, \lambda_2}.
\end{equation*}
Setting $\text{pen}=2$ or $\log(n)$  respectively leads to AIC or BIC.
For the practical evaluation of \eqref{eq:pen_crit}, we must give some
sense      to      the      effective     degrees      of      freedom
$\text{df}_{\lambda_1,\lambda_2}$.    We   use   the   now   classical
definition  of   \cite{2004_JASA_Efron}  and  rely  on   the  work  of
\cite{2012_EJS_Tibshirani} to derive  the following Proposition, the
proof of which is postponed to Appendix Section \ref{sec:df_spring}.

\begin{proposition}\label{prop:df_spring}  An  unbiased  estimator  of
  $\text{df}_{\lambda_1,\lambda_2}$ for our fitting procedure is
  \begin{multline*}
    \hat{\mathrm{df}}_{\lambda_1,  \lambda_2} = \mathrm{card}(\mathcal{A})
    \\ - \lambda_2 \mathrm{tr}\left( (\hat{\bR}\otimes
      \bL )_{\mathcal{A}\mathcal{A}} \left((\hat{\bR}\otimes (\bS_{\bx\bx} +
      \lambda_2\bL))_{\mathcal{A}\mathcal{A}}\right)^{-1}\right),
  \end{multline*}
  where               $\mathcal{A}              =              \set{j:
    \vec\left(\hat{\bOmega}_{\bx\by}^{\lambda_1,\lambda_2}\right) \neq
    0}$     is      the     set     of      nonzero     entries     in
  $\hat{\bOmega}_{\bx\by}^{\lambda_1,\lambda_2}$.
\end{proposition}

Note  that we  can  compute  this expression  at  no additional  cost,
relying on computations already made during the optimization process.


\section{Simulation studies}
\label{Sec:simu}

In this section,  we would like to illustrate the  new features of our
proposal compared  to several  baselines in well  controlled settings.
To this  end, we perform two  simulation studies to  evaluate $i)$ the
gain brought  by the estimation  of the residual covariance  $\bR$ and
$ii)$ the  gain brought by  the inclusion of informative  structure on
the predictors via $\bL$.


\subsubsection*{Implementation    details}    In   our    experiments,
performance are compared with well-established regularization methods,
whose    implementation    is    easily    accessible:    the    LASSO
(\cite{1996_JRSSB_tibshirani}),   the   multitask  group-LASSO,   MRCE
(\cite{2010_JCGS_rothman}), the Elastic-Net (\cite{2005_JRSS_Zou}) and
the  Structured  Elastic-Net  (\cite{2010_AOAS_slawski}).   LASSO  and
group-LASSO  are fitted  with  the \texttt{R}-package  \texttt{glmnet}
(\cite{2010_JSS_friedman})  and  MRCE with  \cite{2010_JCGS_rothman}'s
package.  All  other methods are fitted  using our own code.   Our own
procedure    is   available    as    an   \texttt{R}-package    called
\textbf{spring},             distributed            on             the
\texttt{R-forge}\footnote{\url{https://r-forge.r-project.org/projects/spring-pkg/}}. As
such, we sometimes  refer to our method as 'SPRING'  in the simulation
part.

\subsubsection*{Data   generation}  Artificial  datasets   are  generated
according      to      the      multivariate     regression      model
\eqref{eq:reg_model_row}.   We assume  that the  decomposition  $\bB =
\bOmega_{\bx\by}  \bOmega_{\by\by}^{-1}  = \bOmega_{\bx\by}\bR$  holds
for the  regression coefficients.  We control the  sparsity pattern of
$\bOmega_{\bx\by}$  by  arranging  non  null entries  according  to  a
possible   structure   of   the   predictors   along   the   rows   of
$\bOmega_{\bx\by}$.   We always  use uncorrelated  Gaussian predictors
$\bx_i\sim\mathcal{N}(\bzr,\bI)$ in order not  to promote excessively  the methods
that take  this structure into account. Strength  of the relationships
between  the outputs  are tuned  by the  covariance matrix  $\bR$.  We
measure  the performance  of  the learning  procedures  thanks to  the
prediction error (PE) estimated using a large test set of observations
generated according  to the true  model.  When relevant,  mean squared
error (MSE)  of the regression  coefficients $\bB$ is  also presented.
For conciseness,  it is eluded when it  shows results which are quantitatively 
identical to PE.

\subsection{Influence of covariance between outcomes}

\begin{figure}[htbp!]
  \centering
  \includegraphics[trim=0 0 0 15pt,clip=true,width=.9\columnwidth]{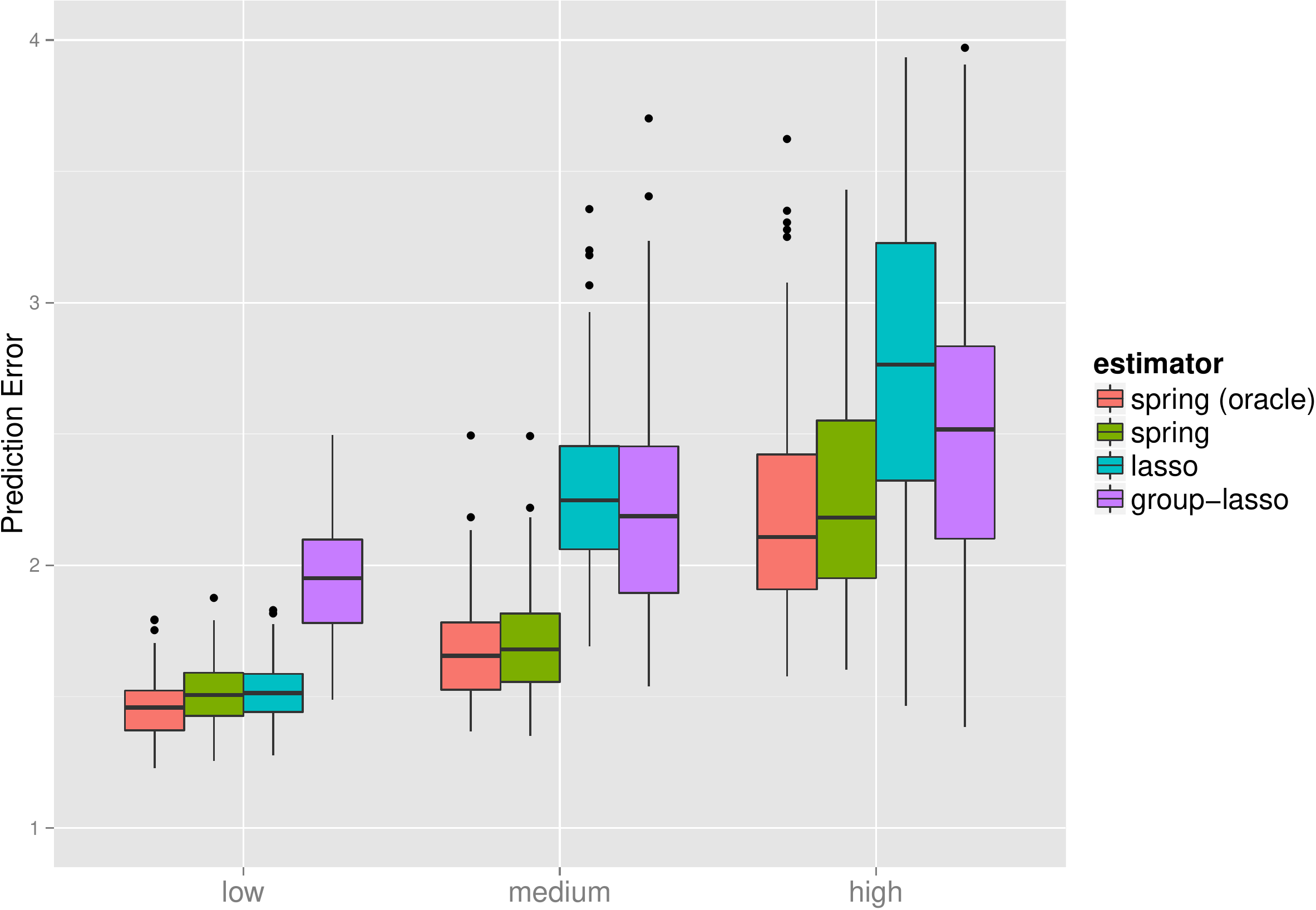}
  \caption{Illustration  of  the  influence  of  correlations  between
    outcomes.  Scenarios $\set{\text{low},\text{med},\text{high}}$ map
    to  $\tau\in\set{.1,.5,.9}$   in  the  Toeplitz-shaped  covariance
    matrix $\bR$.}
    \label{fig:cov_mse}
    \vspace{-.5cm}
\end{figure}

The  first  simulation  study  aims  to illustrate  the  advantage  of
splitting $\bB$  into $-\bOmega_{\bx\by}\bR$ over  working directly on
$\bB$.  We  set $p=40$ predictors, $q=5$ outcomes  and randomly select
$25$ non  null entries in $\set{-1,1}$ to  fill $\bOmega_{\bx\by}$. We
do not put any structure along the predictors as this study intends to
measure the gain of using the cGGM approach.  The covariance follows a
Toeplitz  scheme\footnote{We set  $\bR$ a  correlation matrix  in order not to
  excessively penalize the LASSO or the group-LASSO, which both use the same
  tuning parameter $\lambda_1$ across  the outcomes (and thus the same
  variance  estimator).}: one has      $\bR_{ij}   =  \tau^{|i-j|}$,  for
$i,j=1,\dots,q$.    We  consider  three   scenarios  tuned   by  $\tau
\in\set{.1,.5,.9}$ corresponding to  an increasing correlation between
the  outcomes that  eventually  makes the  cGGM  more relevant.   These
settings    have   been    used   to    generate   panel    $(a)$   of
Figure~\ref{fig:model_insight}.  For   each  covariance  scenario,  we
generate a  training set  with size  $n=50$ and a  test set  with size
$n=1000$.   We assess  the performance  of SPRING  by  comparison with
three baselines:  $i)$ the LASSO, $ii)$  the $\ell_1/\ell_2$ multitask
group-LASSO and $iii)$ SPRING  with known covariance matrix $\bR$.  As
it corresponds to the best fit we can obtain with our proposal, we
call this variant the ``oracle'' mode of SPRING.  The final estimators
are obtained  by 5-fold cross-validation on the  training set.  Figure
\ref{fig:cov_mse}  gives   the  boxplots   of  PE  obtained   for  100
replicates.  As expected, the  advantage of taking the covariance into
account  becomes more  and more  important for  maintaining a  low  PE when
$\tau$ increases.   When correlation is  low, the LASSO  dominates the
group-LASSO;  this is  the other  way  around in  the high  correlation
setup, where  the latter takes  full advantage of its  grouping effect
along the  outcomes. Still, our proposal  remains significantly better
as  soon as  $ \tau$  is  substantial enough.  We also  note that  our
iterative algorithm does a good  job since SPRING remains close to its
oracle variant.

\subsection{Structure   integration   and   robustness}   The   second
simulation study is  designed to measure the impact  of introducing an
informative prior along the predictors via $\bL$. To remove any effect
induced by the covariance between  the outputs, we set $q=1$.  In this
case      the     criterion    is  written      as     in      Expression
\eqref{eq:criteria2_univariate}:   $\bR$  boils   down  to   a  scalar
$\sigma^2$  and $\bB,\bOmega_{\bx\by}$  turn to  two  $p$-size vectors
sharing the same sparsity pattern  such that $\bbeta = -\bomega \sigma^2$.
In  this  situation,  SPRING  is close  to  \cite{2010_AOAS_slawski}'s
structured Elastic-Net, except that we hope for a better estimation of
the  coefficients   thanks  to   the  estimation  of   $\sigma$.   For
comparison,  we thus  draw  inspiration from  the simulation  settings
originally  used  to illustrate  the  structured  Elastic-Net: we  set
$\bomega  = (\omega_j)_{j=1}^p,  $ with  $p=100$, so  that we  observe a
sparse vector  with two smooth bumps,  one positive and  the other one
negative:
\begin{equation*}
  \omega_j =
  \begin{cases}
    -( (30-j)^2 -100)/200 & j=21,\dots 39,\\
    \ \ \ ( (70-j)^2 -100)/200 & j=61,\dots 80,\\
    0 & \text{otherwise}.
  \end{cases}
\end{equation*}
A natural choice for $\bL$ is  to rely on the first forward difference
operator  as in the  fused-Lasso (\cite{2005_JRSSB_tibshirani}),  or its
smooth counterpart (\cite{2011_EJS_Hebiri}).  We thus set 
$\bL = \bD^T\bD$ with
\begin{equation}
  \label{eq:first_order_diff}
  \bD_{ij} = \begin{cases}
    1 & \text{if } i = j,\\
    -1 & \text{if } j = i+1 \\
    0 & \text{otherwise}\\
  \end{cases},
  \quad \begin{array}{c}
    i=1,\dots,p-1, \\
    j=1,\dots,p.
  \end{array}
\end{equation}
Hence, $\bL$ is  the combinatorial Laplacian of a  chain graph between
the successive predictors.
\begin{figure}[htpb!]
  \centering
  \includegraphics[trim=0 0 0 15pt,clip=true,width=.9\columnwidth]{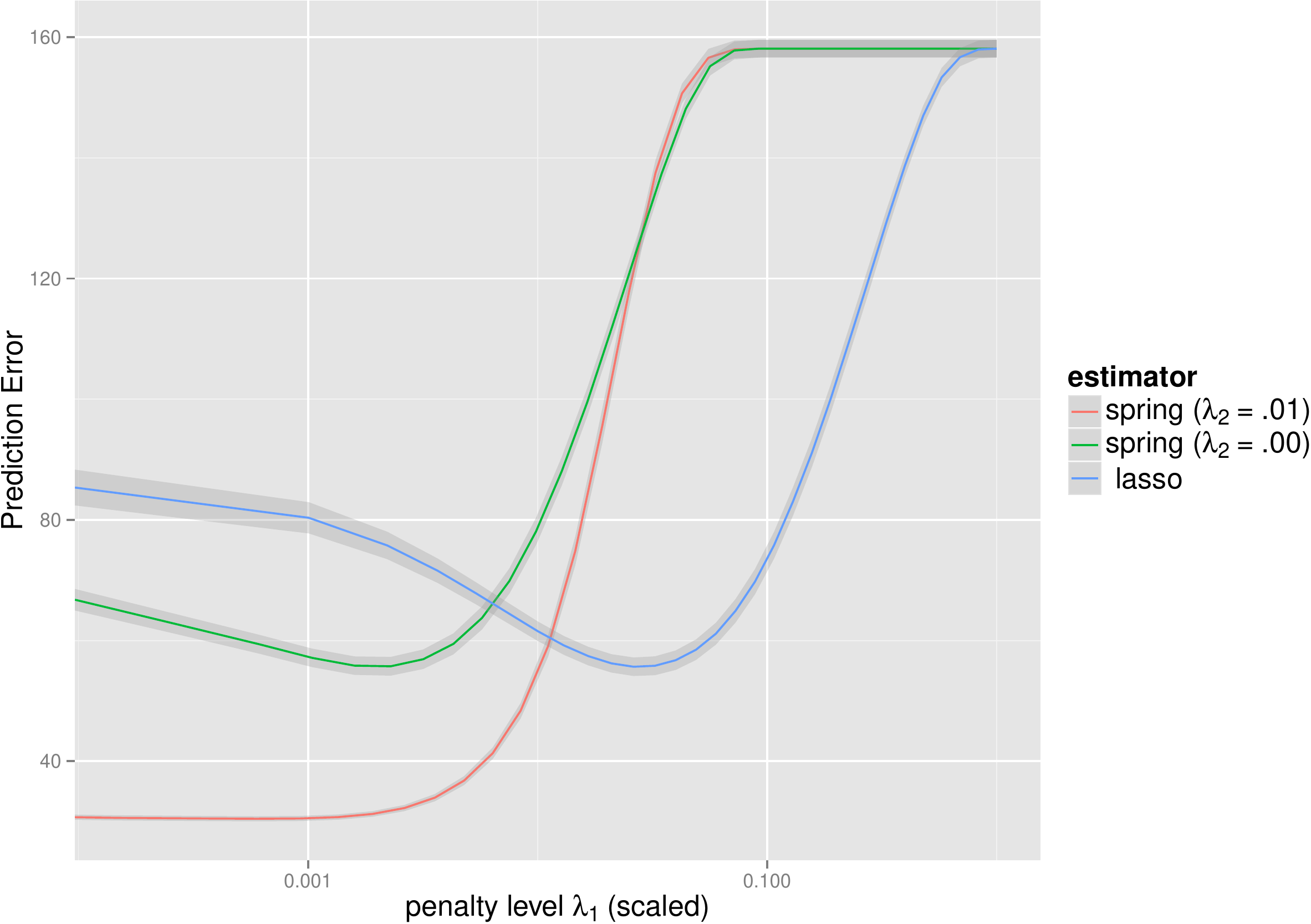}
  \caption{Illustrating gain  brought by structure  information toward
    predictive performance}
  \label{fig:struct_illustration}
\end{figure}
Figure~\ref{fig:struct_illustration} shows the typical gain brought by
prior structure knowledge.  We generate $100$ learning samples of size $n=100$
with $\sigma^2=5$ and  represent the averaged PE curves  as a function
of  $\lambda_1$ for the  LASSO and  for two  versions of  SPRING, with
($\lambda_2 =  .01$) and without ($\lambda_2 =  0$) informative prior.
Incorporating relevant structural  information leads to  a dramatic
improvement.  As  expected, univariate  SPRING with no  prior performs
like the LASSO, in the sense that they share the same minimal PE.

Now, the question is:  what if we introduce a wrong  prior, i.e. a
matrix $\bL$  completely disconnected from  the true structure  of the
coefficients?  To  answer this question,  we use the same  settings as
above and  randomly swap the  entries in $\bomega$, using  exactly the
same $\bx_i$  to generate the  $\by_i$\footnote{We also used  the same
  seed and CV-folds for both  scenarios.}.  We then apply SPRING using
respectively  a non  informative prior  equal to  the identity  matrix
(thus mimicking the Elastic-Net) and  a 'wrong' prior $\bL$ whose rows
and columns remain unswapped.  We  also try the LASSO, the Elastic-Net
and the  structured Elastic-Net.   All methods  are tuned  with 5-fold
cross-validation  on  the  learning  set.   Table~\ref{tab:struct_res}
presents the results averaged over 100  runs both in terms of PE (using
a size-1000 test set) and MSE.
\begin{table}[htbp!]
  \centering
  \begin{small}
    \begin{tabular}{@{}lrrr@{}}
    \hline
    \textit{\textsf{Method}} & \textit{\textsf{Scenario}} & \textit{\textsf{MSE}} & \textit{\textsf{PE}} \\
    \hline
    LASSO
    & -- & .336 (.096) & 58.6 (10.2) \\
    \cline{1-4}
    E-Net    ($\bL=\bI$)
    & -- & .340 (.095) & 59 (10.3) \\
    \cline{1-4}
    SPRING    ($\bL=\bI$)
    & -- & .358 (.094) & 60.7 (10) \\
    \cline{1-4}
    S. E-net
    & unswapped & .163 (.036) & 41.3 ( 4.08) \\
    ($\bL=\bD^T \bD$)
    & swapped & .352 (.107) & 60.3 (11.42) \\
    \cline{1-4}
    SPRING
    & unswapped & .062 (.022) & 31.4 ( 2.99) \\
    ($\bL=\bD^T \bD$)
    & swapped & .378 (.123) & 62.9 (13.15) \\
    \hline
  \end{tabular}
  \end{small}
  \caption{Structure integration: performance and robustness. Swapped and unswapped results are the same for LASSO, Elastic-Net and SPRING for $\bL = \bI$.}
  \label{tab:struct_res}
\end{table}

As  expected,  the  methods  that  do  not  integrate  any  structural
information  (LASSO, Elastic-Net  and SPRING  with $\bL=\bI$)  are not
affected by the permutation, {and we avoid these redundancies in Table
  \ref{tab:struct_res} to  save space}.   Overall, they  share similar
performance both in  terms of PE and MSE.  When  the prior structure is
relevant, SPRING, and  to a lesser extent  the structured Elastic-Net,
clearly  outperform  the  other competitors.   Surprisingly,  this  is
particularly  true in  terms of  MSE, where  SPRING also  dominates the
structured Elastic-Net  that works with the same  information.  This
means that the estimation of the variance also helped in the inference
process.  Finally, these results essentially support the robustness of
the structured methods  which are not much altered when  using a wrong
prior specification.


\section{Application Studies}
\label{Sec:applications}

In  this section  the flexibility  of our  proposal is  illustrated by
investigating three multivariate  data problems from various contexts,
namely  spectroscopy, genetics and  genomics, where  we insist  on the
construction of the structuring matrix $\bL$.
%
%

\subsection{Near-Infrared Spectroscopy of Cookie Dough Pieces}

\subsubsection*{Context} In Near-Infrared (NIR)  spectroscopy, one aims to
predict one or several quantitative variables from the NIR spectrum of
a given sample.  Each sampled spectrum  is a curve that represents the
level of reflectance  along the NIR region, that  is, wavelengths from
800 to 2500 nanometers (nm).  The quantitative variables are typically
related to  the chemical  composition of the  sample.  The  problem is
then to  select the most predictive  region of the spectrum,  i.e.
some peaks  that show  good capabilities  for predicting  the response
variable(s).   This   is  known   as  a  ``calibration   problem''  in
Statistics.  NIR technique  is used in fields as  diverse as agronomy,
astronomy or pharmacology.  In such  experiments, it is likely to encounter
very strong correlations and structure  along the predictors.  In this
perspective, \cite{2011_JASA_hans}  proposes to apply  the Elastic-Net
which  is  known  to  select  simultaneously  groups  of  correlated
predictors. However  it is  not adapted to  the prediction  of several
responses  simultaneously.  In  \cite{data_cookies2}, an  interesting
wavelet regression  model with Bayesian inference  is introduced that
enters the multivariate regression model, as does our proposal.

\subsubsection*{Description of  the dataset} We consider  the cookie dough
data from \cite{data_cookies1}.  The  data with the corresponding test
and  training  sets  are  available  in  the  \textbf{fds}  \texttt{R}
package.  After data pretreatments as in \cite{data_cookies2}, we have
$n=39$ dough  pieces in the training  set: each sample  consists in an
NIR spectrum with $p=256$ points measured from 1380 to 2400 nm (spaced
by  4  nm), and  in  four  quantitative  variables that  describe  the
percentages of fat, sugar flour and water of in the piece of dough.

\subsubsection*{Structure specification} We would  like to account for the
neighborhood structure between the  predictors which is obvious in the
context of  NIR spectroscopy: since spectra are  continuous curves, a
smooth neighborhood prior will  encourage predictors to be selected by
``wave'', which seems more  satisfactory than isolated peaks. Thus, we
naturally  define  $\bL$ by  means  of  the  first forward  difference
operator \eqref{eq:first_order_diff}.  We also tested higher orders of
the operator  to induce a stronger  smoothing effect, but  they do not
lead to dramatic changes in terms of PE and we omit them.  Order $k$ is
simply obtained  by powering the first order  matrix.  Such techniques
have  been  studied  in  a structured  $\ell_1$-penalty  framework  in
\cite{2009_SIAM_Kim,2011_AOS_Tibshirani}  and is known  as \emph{trend
  filtering}.   Our  approach however is   different,  though:  it  enters  a
multivariate framework  and is  based on a  smooth $\ell_2$-penalty
coupled with the $\ell_1$-penalty for selection.

\subsubsection*{Results}  
\begin{figure*}[htbp!]
  \centering
  \begin{tabular}{@{}c@{}c@{}c@{}c}
    &  $\hat{\bB}$   (indirect  effects)  &  $-\hat{\bOmega}_{\bx\by}$ (direct effects) & $\hat{\bR}$ \\
    \includegraphics[width=.08\textwidth,clip=true,trim=0 -5cm 0 0]{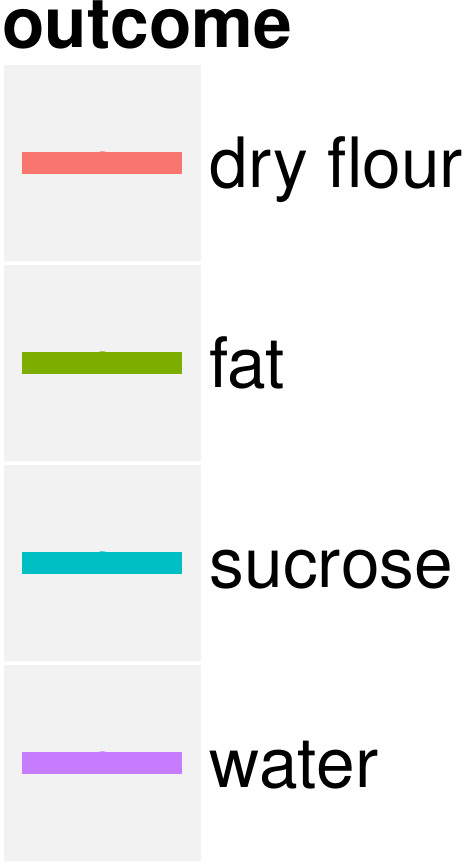}
    & \includegraphics[width=.3\textwidth]{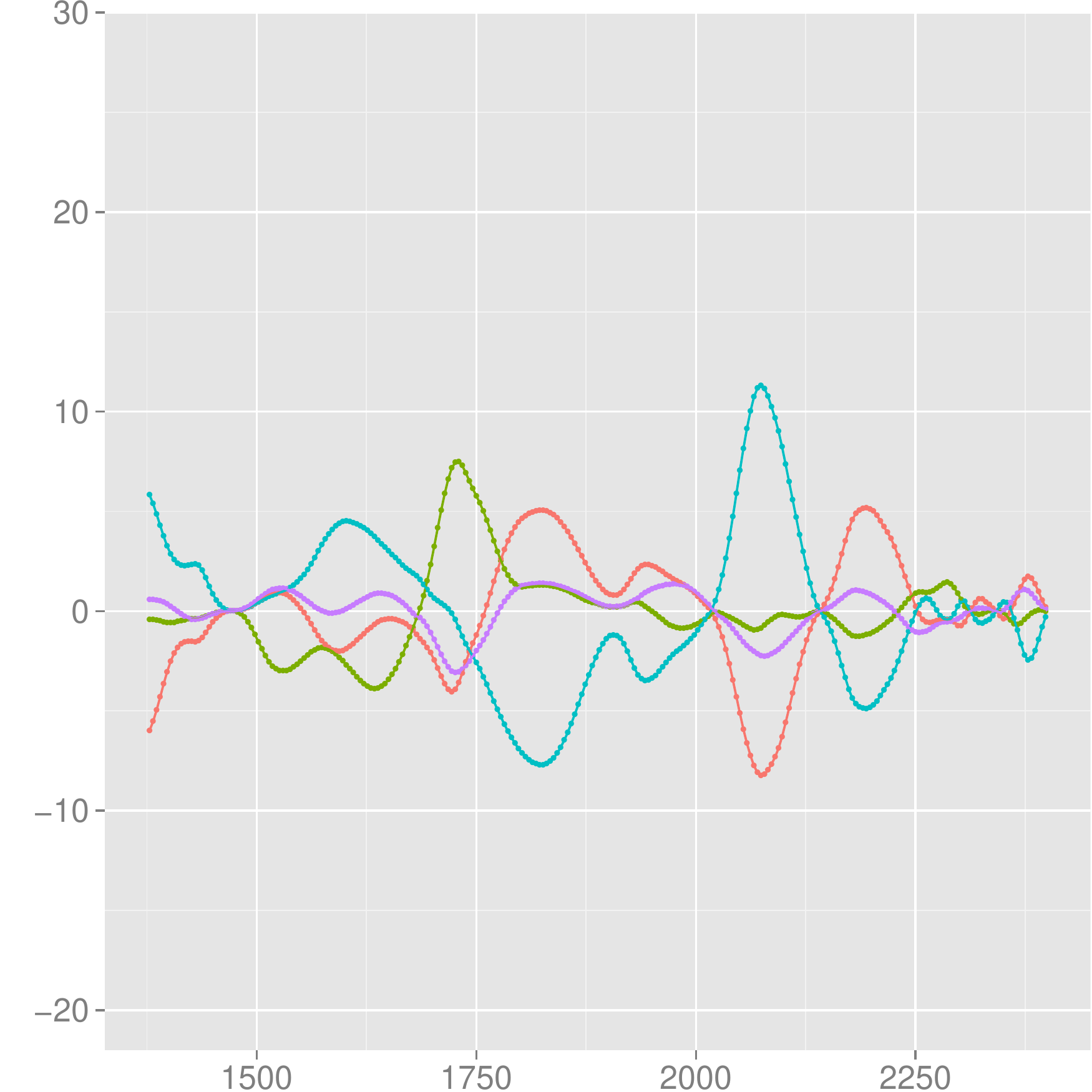}
    & \includegraphics[width=.3\textwidth]{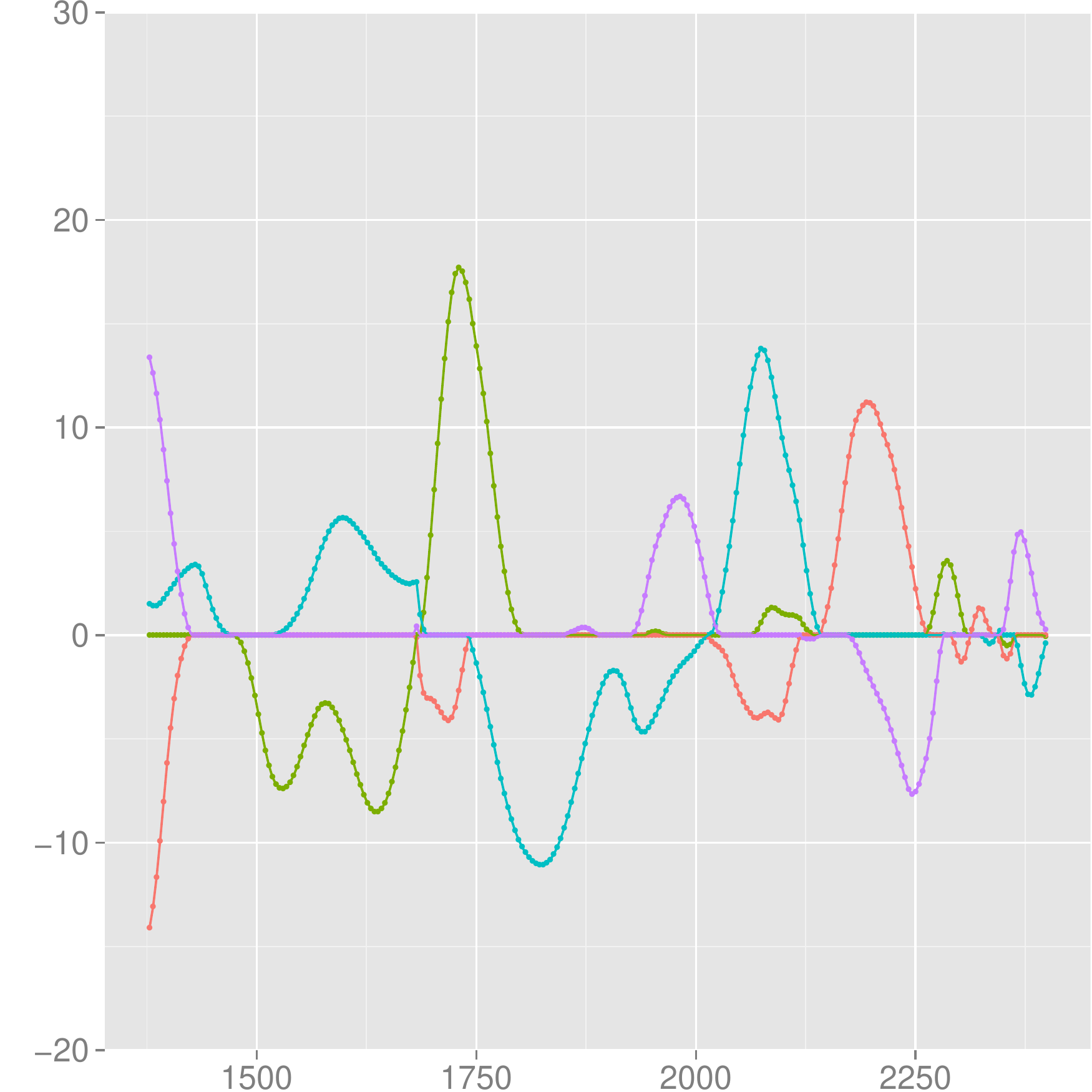}
    & \includegraphics[width=.3\textwidth]{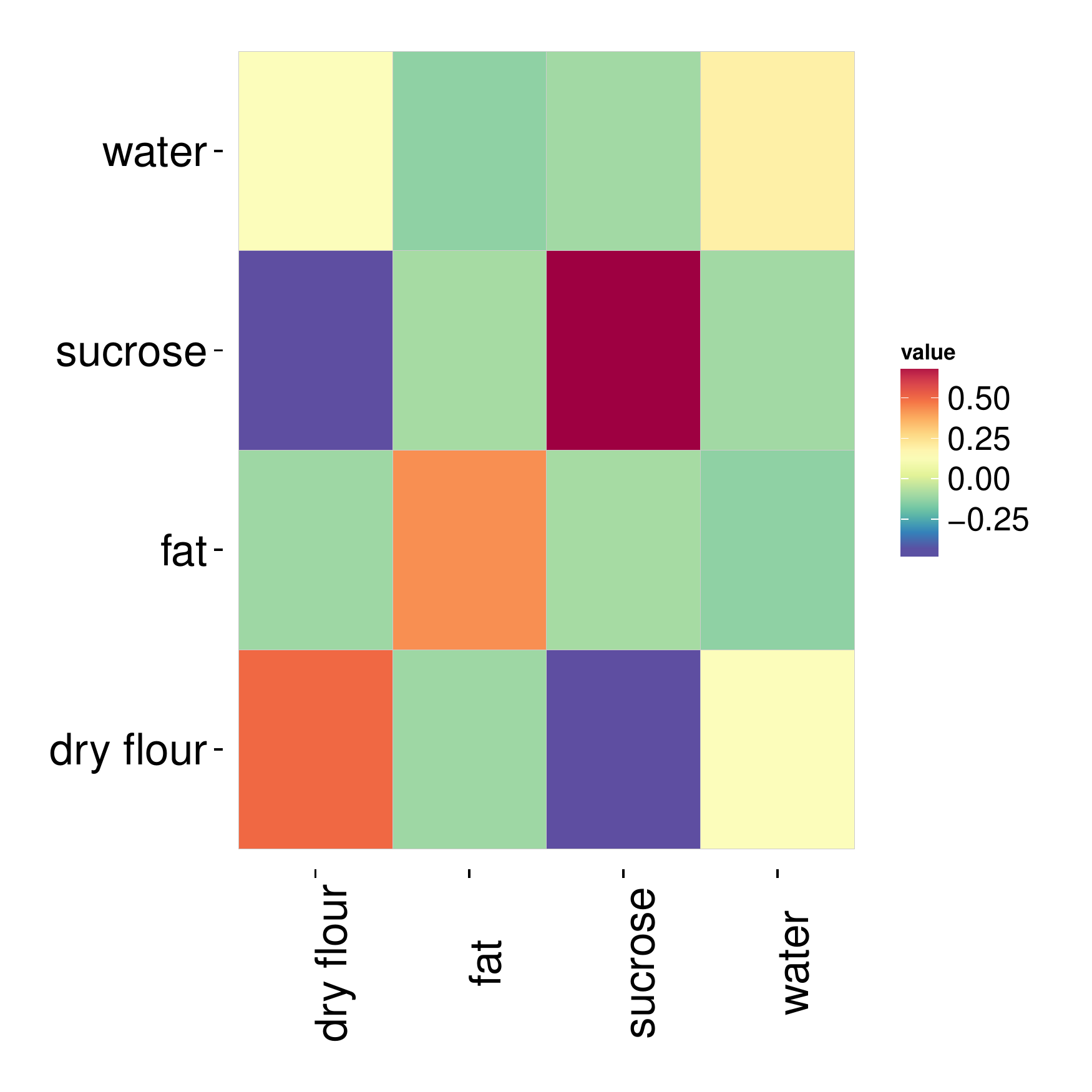}\\
    & \multicolumn{2}{c}{wavelength range (nanometer)} & \\
  \end{tabular}
  \caption{Parameters estimated by our  proposal for the cookie dough
    data   (better   seen   in   color).    {Since   $\hat{\bB}$   and
      $\hat{\bOmega}_{\bx\by}$ are  opposite in sign in  our model, we
      represent  $-\hat{\bOmega}_{\bx\by}$   to  ease  the  comparison
      between direct and indirect effects.}}
  \label{Fig:Cookies1}
\end{figure*}

The  predictive  performance  of  a  series  of
regression  techniques   is  compared  in   Table  \ref{Tab:Cookies},
evaluated on the same test set.
\begin{table}[htbp!]
  \centering
  \begin{small}
    \begin{tabular}{l|rrrr}
      \hline
      \textit{\textsf{Method}} & \textit{\textsf{fat}} & \textit{\textsf{sucrose}} & \textit{\textsf{flour}} & \textit{\textsf{water}} \\
      \hline
      {Stepwise MLR} & \textbf{.044} & 1.188 & .722 & .221 \\
      {Decision theory} & .076 & .566 & .265 & .176 \\
      {PLS} & .151 & .583 & .375 & .105 \\
      {PCR} & .160 & .614 & .388 & .106 \\
      {Wavelet Regression} & .058 & .819 & .457 & .080 \\
      {LASSO} & \textbf{.044} & .853 & .370 & .088 \\
      {group-LASSO} & .127 & .918 & .467 & .102 \\
      {MRCE} & .151 & .821 & .321 & .081 \\
      {Structured E-net} & \textbf{.044} & .596 & .363 & .082 \\
      {SPRING} (CV) & .065 & .397 & \textbf{.237} & .083 \\
      {SPRING} (BIC)& .048 & \textbf{.389} & .243 & \textbf{.066} \\
  \end{tabular}
  \end{small}
  \caption{Prediction error for the cookie dough data.}
  \label{Tab:Cookies}
\end{table}
The first four rows (namely stepwise multivariate regression, Bayesian
decision theory approach, Partial Least Square and Principal Component
Regression) correspond to  calibration methods originally presented in
\cite{data_cookies2};  row  five  (Bayesian  Wavelet  Regression)  is
the original proposal from \cite{data_cookies2}; the remainder of the
table is due  to our own analysis.  We observe  that, on this example,
SPRING  achieves  extremely  competitive  results.  The  BIC  performs
notably well as a model selection criterion.
\begin{figure*}[htbp!]
  \centering
  \begin{tabular}{@{}c@{}c@{}c@{}c@{}}
    LASSO & group-LASSO &  Structured Elastic-Net & MRCE \\
    \includegraphics[width=.25\textwidth]{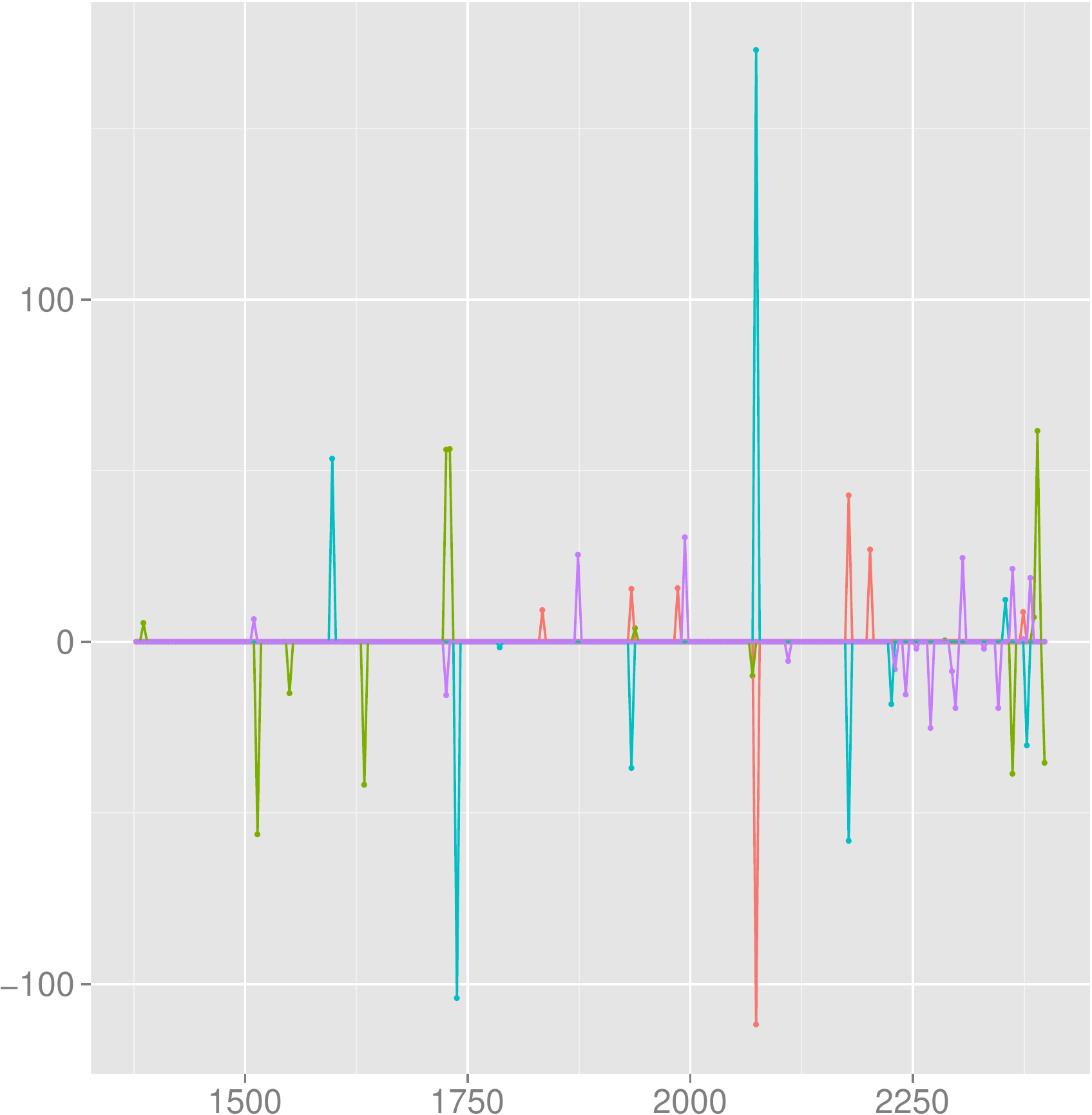} &
    \includegraphics[width=.25\textwidth]{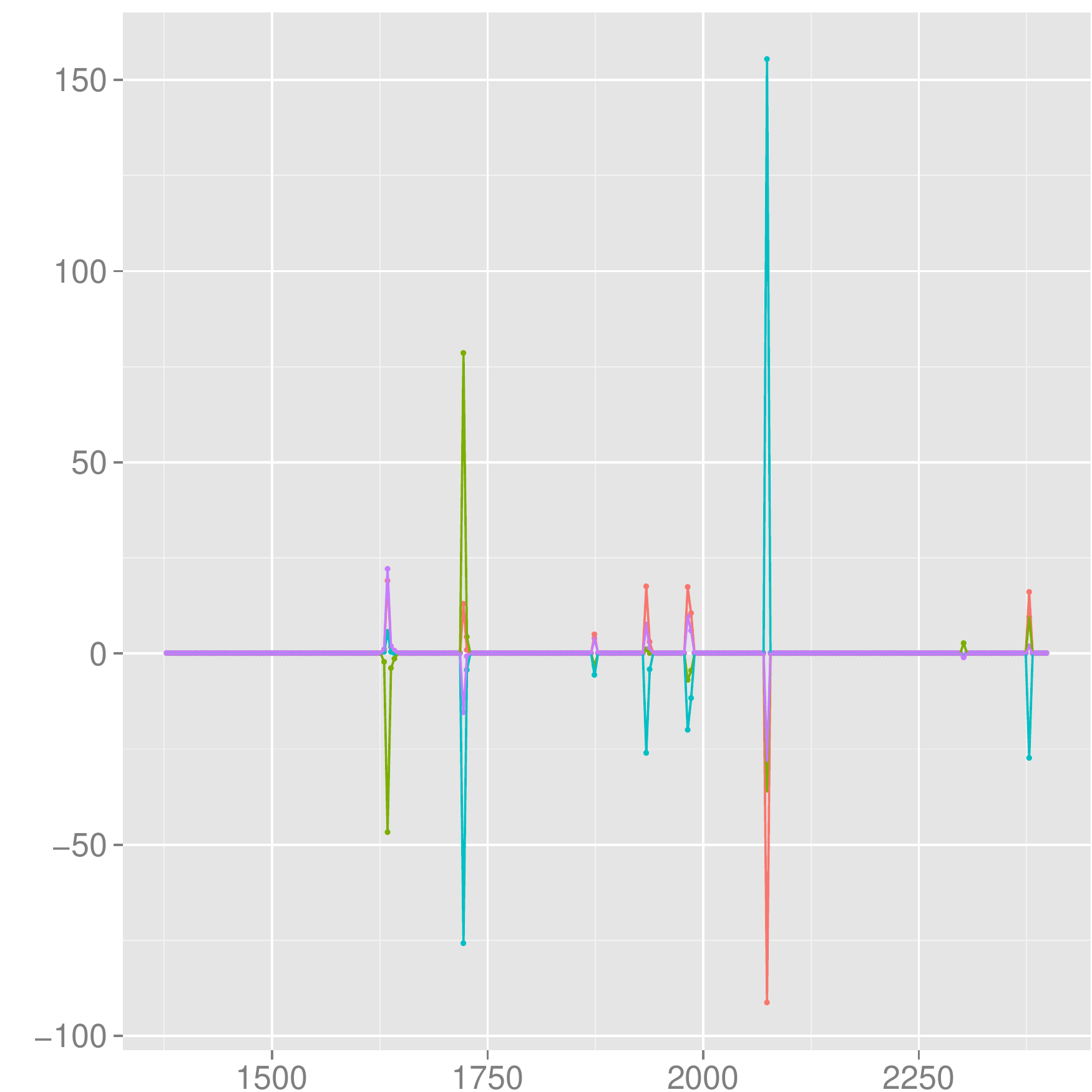} &
    \includegraphics[width=.25\textwidth]{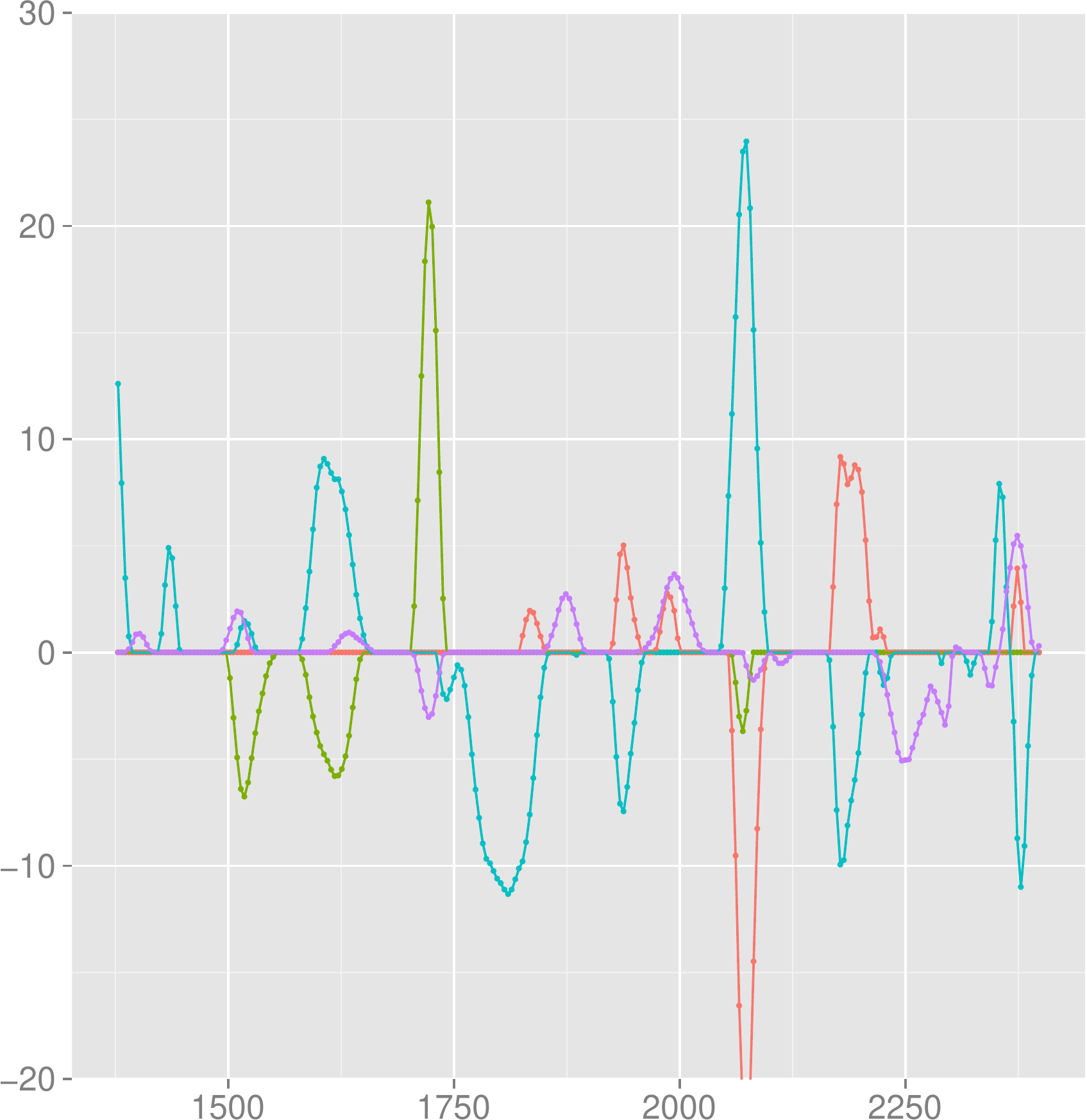} &
    \includegraphics[width=.25\textwidth]{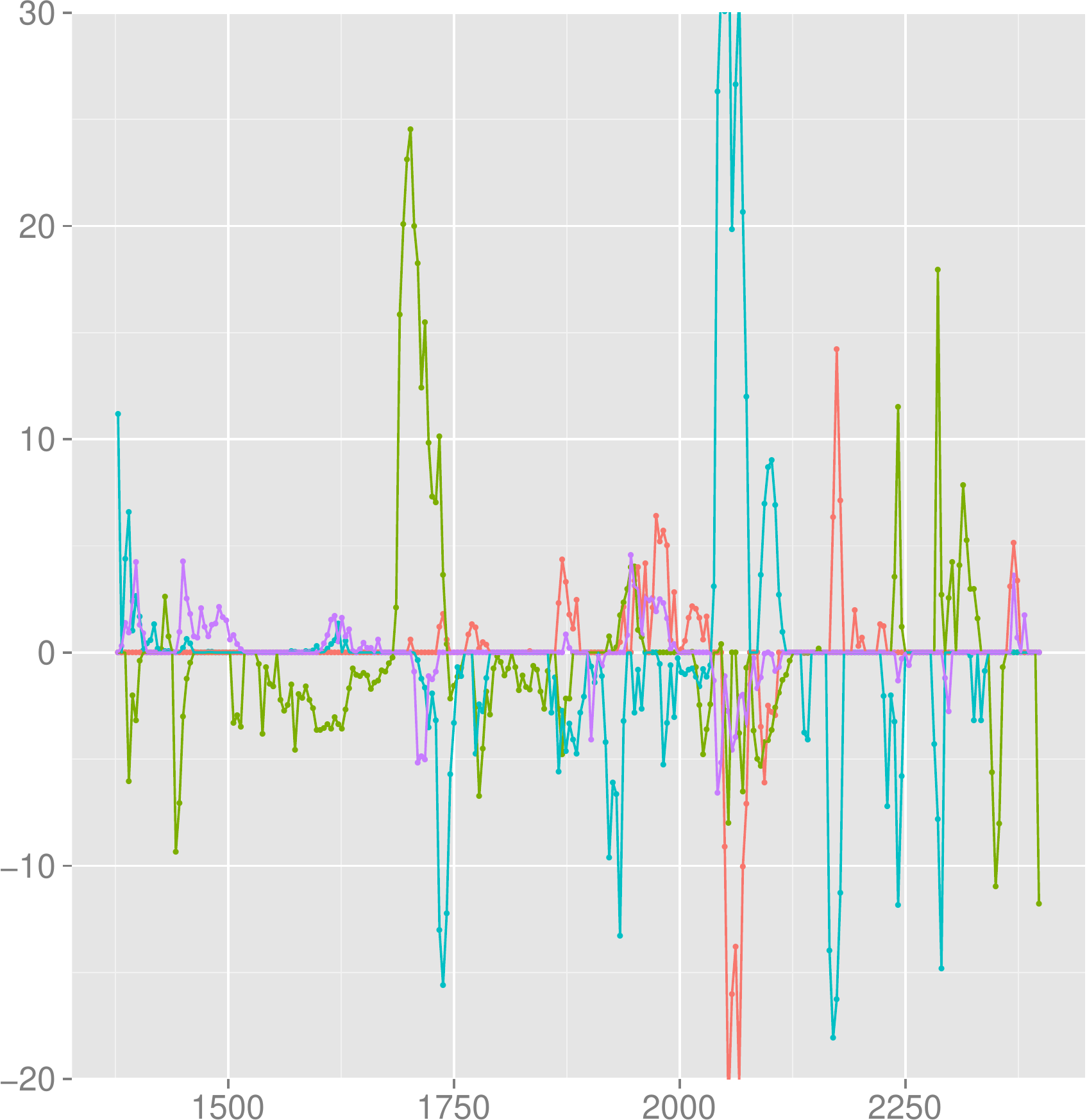} \\
    \multicolumn{4}{c}{Wavelength range (nanometer)}
  \end{tabular}
  \caption{Parameters estimated by the penalized regression methods for
    the cookie data}
  \label{Fig:Cookies2}
\end{figure*}
Figure \ref{Fig:Cookies2}  shows the regression  coefficients adjusted
with  the  penalized regression  techniques:  apart  from LASSO  which
selects very isolated  (and unstable) wavelengths, non-zero regression
coefficients  have quite a wide spread,   and therefore  hard to  interpret.  As
expected,  the  multitask group-Lasso  activates  the same  predictors
across the  responses, which is  not a good  idea when looking  at the
predictive performance in Table \ref{Tab:Cookies}.  On the other hand,
in Figure~\ref{Fig:Cookies1}, the
direct effects  selected by SPRING with BIC  define predictive regions
specific to  each response which are well  suited for interpretability
purposes.
Parameters  fitted  by  our  approach  with  BIC  are  represented  on
Figure~\ref{Fig:Cookies1} with, from left  to right, the estimators of
the   regression   coefficients   $\bB$,   of   the   direct   effects
$\bOmega_{\bx\by}$  and   of  the  residual   covariance  $\bR$.   The
regression coefficients show no  sparsity pattern but a strong spatial
structure  along the  spectrum, characterized  by waves  induced  by a
smooth first order difference prior.  Concerning a potential structure
between the  outputs, we identify  interesting regions where  a strong
correlation  between the  responses induces  correlations  between the
regression  coefficients.   Consider  for  instance positions  of  the
wavelength between 1750 and 2000 nm: the regression parameters related
to  ``dry-flour''  are clearly  anti-correlated  with  those related  to
``sucrose''.  Still,  we cannot distinguish in $\hat{\bB}$  for a direct
effect  of this  region on  either the  flour or  sucrose composition.
Such  a distinction  is  achieved  on the  middle  panel where  direct
effects  $\hat{\bOmega}_{\bx\by}$  selected  are plotted:  it  defines
sparse  predictive regions specific  to each  response which  are well
suited for  interpretability purposes; in fact,  it is now
obvious that region 1750 to 2000 is rather linked to the sucrose.


\subsection{Multi-trait Genomic Selection in Brassica napus}

\subsubsection*{Context} Genomic selection is aimed at predicting  one or several phenotypes
based on the information of genetic markers.
To this end, regularization methods  such as ridge or Lasso regression
or     their    Bayesian     counterparts    have     been    proposed
(\cite{LosCampos12}).   Still,  in  most  studies  only  single  trait
genomic  selection  is   performed,  neglecting  correlations  between
phenotypes
.   Moreover,  little attention  has  been  devoted to  the
development   of  regularization   methods  including   prior  genetic
knowledge.

\subsubsection*{Description of the dataset} We consider the \emph{Brassica
  napus}     dataset    described    in     \cite{Bnapus_data}    and
\cite{Bnapus_map}.   Data consists  in  $n=103$ double-haploid  lines
derived from 2 parent cultivars,  `Stellar' and `Major', on which $p =
300$ genetic markers and  $q=8$ traits (responses) were recorded. Each
marker is a 0/1 covariate with $x_i^j=0$ if line $i$ has the 'Stellar'
allele  at marker $j$,  and $x_i^j=1$  otherwise. Traits  included are
percent winter survival  for 1992, 1993, 1994, 1997  and 1999 (surv92,
surv93, surv94,  surv97, surv99, respectively), and  days to flowering
after no vernalization (flower0), 4 weeks vernalization (flower4) or 8
weeks vernalization (flower8).

\subsubsection*{Structure specification}
In a biparental line population, correlation between 2 markers depends
on their genetic distance defined  in terms of recombination fraction.
As a consequence,  one expects adjacent markers on the  sequence to be
correlated, yielding similar direct  relationships with the phenotypic
traits. Noting $d_{12}$ the genetic distance between markers $M_1$ and
$M_2$, one has $\text{cor}(M_1,M_2)  = \rho^{d_{12}}$, where $\rho =
.98$\footnote{This value  directly arises  from the definition  of the
  genetic  distance itself.}.   The covariance  matrix $\bL^{-1}$  can
hence  be  defined  as  $\bL_{ij}^{-1}  =  \rho^{d_{ij}}$.   Moreover,
assuming recombination events are  independent between $M_1$ and $M_2$
on the one hand, and $M_2$ and $M_3$  on the other hand, one has $d_{13} =
d_{12} + d_{23}$ and matrix $\bL^{-1}$ exhibits an inhomogeneous AR(1)
profile.  As a consequence, $\bL$ is tridiagonal with general elements
\begin{eqnarray*}
  w_{i, i} & = & \frac{1 - \rho^{2d_{i-1, i} + 2d_{i, i+1}}}{(1 - \rho^{2d_{i-1, i}})(1 - \rho^{2d_{i, i+1}})}, \\
  w_{i, i+1} & = & \frac{-\rho^{d_{i, i+1}}}{1 - \rho^{2d_{i, i+1}}}
\end{eqnarray*}
and $w_{i,j} =  0$ if $|i-j| > 1$. For the  first (resp. last) marker,
the distance $d_{i-1, i}$ (resp. $d_{i, i+1}$) is infinite.

\subsubsection*{Results} 
\begin{figure}[htpb!]
  \includegraphics[scale=.35]{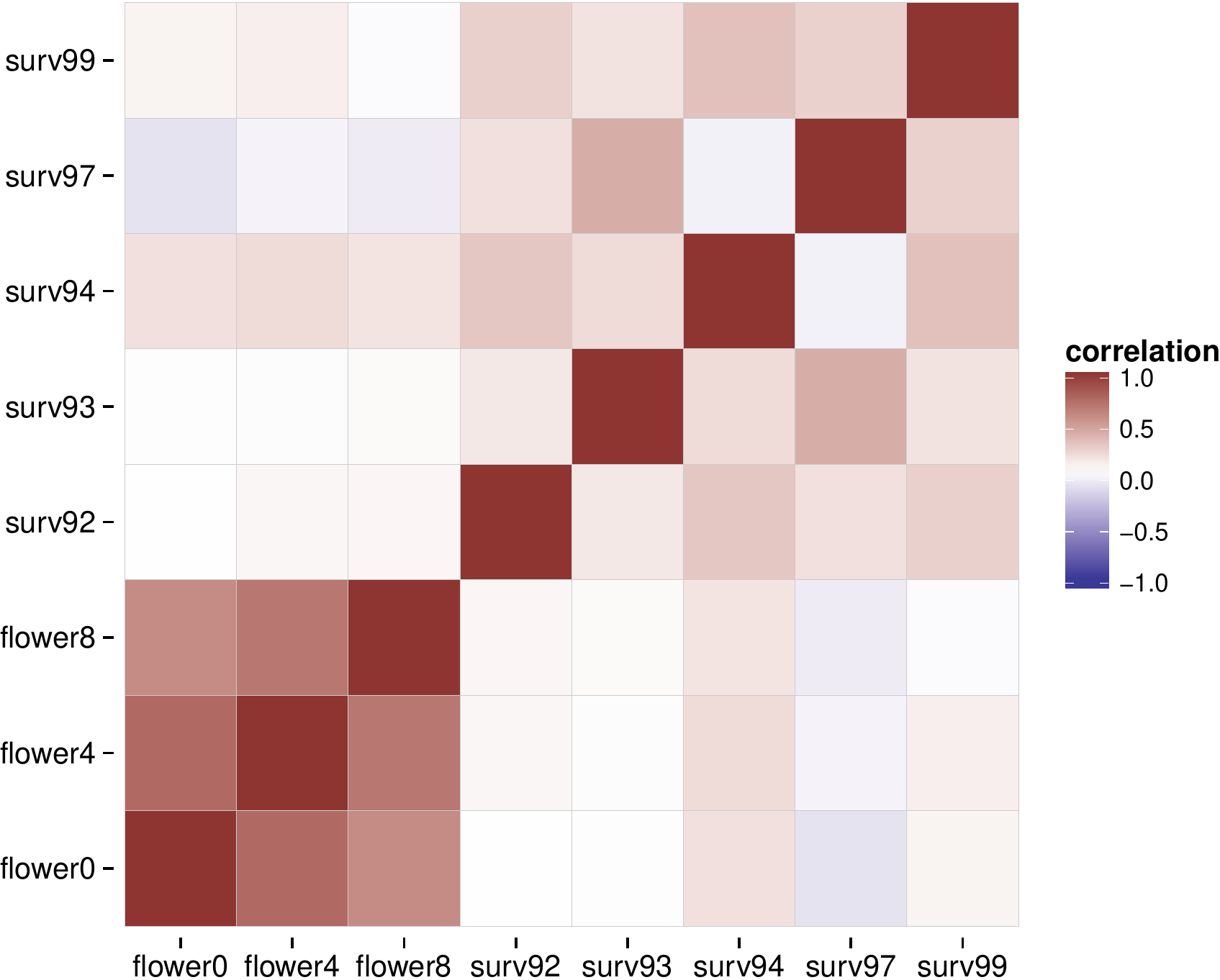}
  \caption{Brassica study: residual covariance estimation \label{Fig:BNapusCovar}}
\end{figure}

To compare SPRING and its competitors in terms of predictive performance,
%
   PE is estimated by randomly  splitting the $103$
samples  into training  and test  sets with  sizes 93  and  10. Before
adjusting the models, we first  scale the outcomes on the training and
test sets  to facilitate interpretability.  Five-fold cross-validation
is  used on the  training set  to choose  the tuning  parameters.  Two
hundred random samplings of the  test and training sets were conducted
to estimate the PE given in Table \ref{Tab:BNapus}.
\begin{table*}[htp!]
  \centering
  \begin{small}
    \begin{tabular}{l|rrrrr}
    \hline
    \textit{\textsf{Method}} &
    \textit{\textsf{surv92}}  &
    \textit{\textsf{surv93}}  &
    \textit{\textsf{surv94}}  &
    \textit{\textsf{surv97}}  &
    \textit{\textsf{surv99}} \\
    \hline
    {LASSO} & .730 (.011) & .977 (.009) & .943 (.010) & .947 (.009) & .916
    (.010) \\

    {S. Enet}  & \textbf{.697} (.011) &  .987 (.009) & .941  (.011) & .945
    (.009) & .911 (.010) \\

    {MRCE} & .759 (.010)  & \textbf{.919} (.003) & 917 (.006) & \textbf{.924} (.004) & .926
    (.006) \\

    {SPRING} &  .724 (.010) &  .948 (.008) & \textbf{.848}  (.010) & .940  (.006) &
    \textbf{.907} (.009) \\
  \end{tabular}
  \medskip

  \begin{tabular}{l|rrr}
    \hline
    \textit{\textsf{Method}} &
    \textit{\textsf{flower0}} &
    \textit{\textsf{flower4}} &
    \textit{\textsf{flower8}} \\
    \hline
    {LASSO} & .609 (.011) & .501 (.011) & .744 (.011) \\

    {S. Enet} & .577 (.011) & .478 (.010) & .727 (.012) \\

    {MRCE} & .591 (.011) & .479 (.011) & .736 (.011) \\

    {SPRING} & \textbf{.489} (.010) & \textbf{.419} (.009) & \textbf{.616} (.012) \\
  \end{tabular}

  \end{small}
  \caption{Estimated  prediction error  for the  \emph{Brassica napus}
    data (standard error in parentheses)}
  \label{Tab:BNapus}
\end{table*}
All  methods  provide similar  results  although  SPRING provides  the
smallest  error   for  half   of  the  traits.    A  picture   of  the
between-response covariance  matrix estimated with SPRING  is given in
Figure \ref{Fig:BNapusCovar}.   It reflects
the correlation  between the  traits, which are  either explained  by an
unexplored  part  of the  genotype,  by  the  environment or  by  some
interaction  between the two.   The residuals  of the  flowering times
exhibit  strong  correlations,  whereas  correlations  between  the
survival rates are weak. It also shows that the survival traits have a
larger residual  variability than do the flowering traits,  suggesting a
higher sensitivity to environmental conditions.

\begin{figure*}[htbp!]
   \centering
   \includegraphics[width=\textwidth]{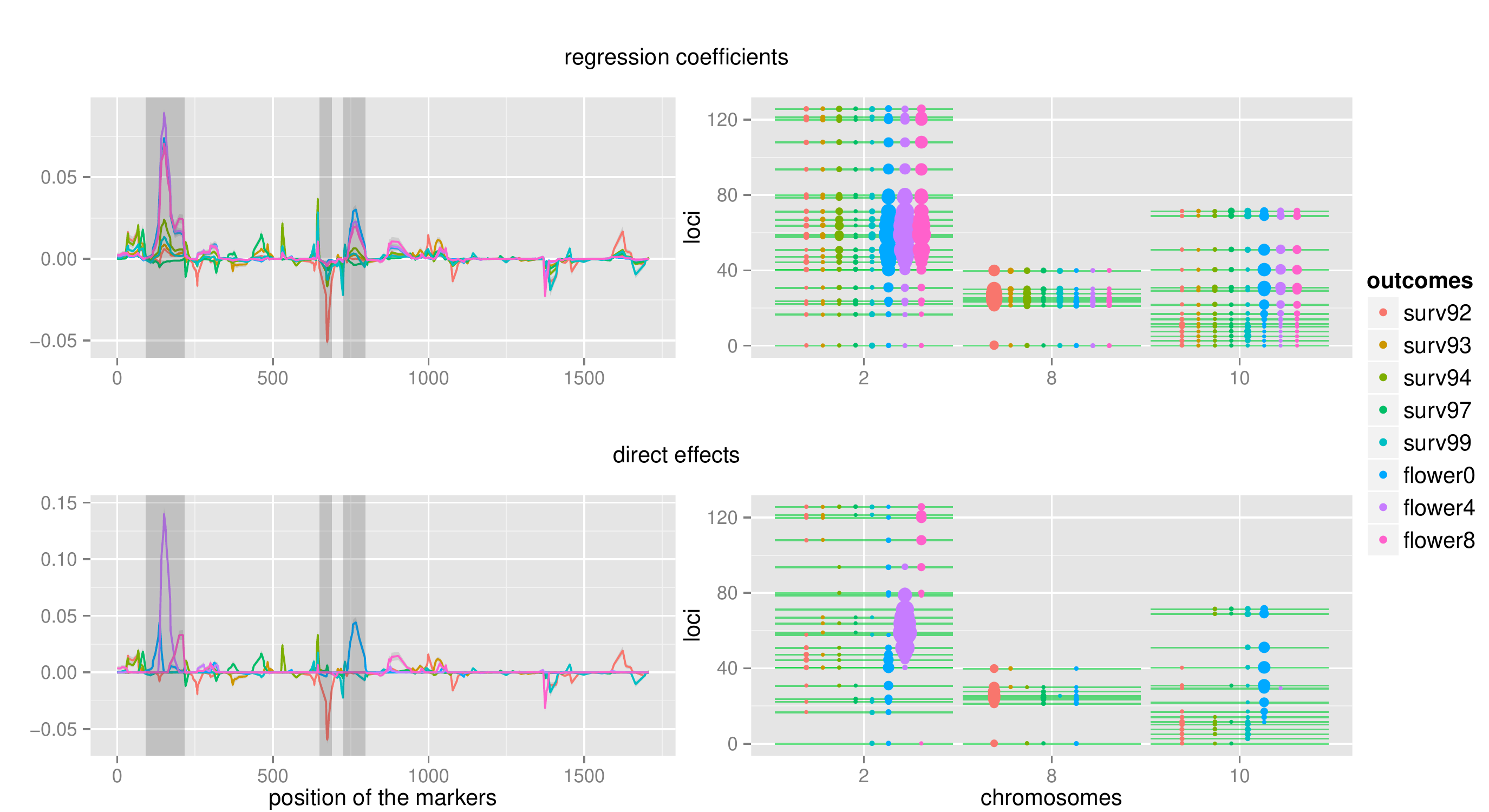}
   \caption{Brassica Study: direct and indirect genetic effects of the
     markers on the  traits estimated by SPRING\label{Fig:BNapusCoefs}
     (better seen in color).}
\end{figure*}

We  then  turn  to  the  effects  of  each  marker  on  the  different
traits. The left panels of Figure \ref{Fig:BNapusCoefs} give both the
regression  coefficients (top)  and the  direct effects  (bottom). The
gray zones correspond  to chromosomes 2, 8 and  10, respectively.  The
exact location  of the markers within these  chromosomes is displayed
in the right panels, where the  size of the dots reflects the absolute
value of the  regression coefficients (top) and of  the direct effects
(bottom).
The  interest of  considering  direct effects  rather than  regression
coefficients appears  clearly here, if one looks for  example at chromosome
2.  Three  large overlapping regions  are observed in  the coefficient
plot, for each flowering trait. A straightforward interpretation would
suggest that  the corresponding region controls  the general flowering
process. The  direct effect  plot allows one to  go deeper and  shows that
these  three  responses are  actually  controlled  by three  separate
sub-regions within  this chromosome. The confusion  in the coefficient
plot only  results from the  strong correlations observed  among the
three flowering traits.


\subsection{Selecting regulatory motifs from multiple microarrays}

\subsubsection{Context} In genomics, the  expression of genes is initiated
by transcription factors that bind to the DNA upstream from the coding
regions, called {\it regulatory  regions}.  This binding occurs when a
given   factor   recognizes   a   certain  (small)   sequence   called
a {\it  regulatory motif}. Genes  hosting the same  regulatory motif will  be jointly
expressed under certain conditions.  As the binding relies on chemical
affinity, some degeneracy can be tolerated in the motif definition, and
motifs similar  but for small  variations may share the  same functional
properties (see, e.g. \cite{2012_GB_Lajoie}).

We  are interested  in the  detection of  such regulatory  motifs, the
presence of  which controls the gene expression profile. To this  aim we
try to establish a relationship between  the expression level of all genes
across a  series of  conditions with the  content of  their respective
regulatory regions in terms of motifs. In this context, we expect $i)$
the set  of influential motifs to  be small for each  condition, $ii)$
the influential motifs  for a given condition to  be degenerate versions
of each other,  and $iii)$ the expression under  similar conditions to
be controlled by the same motifs.

\subsubsection*{Description of the  dataset} In \cite{gasch2000genomic}, a
series  of  microarray  experiments   are  conducted  on  yeast  cells
(\textit{Saccharomyces cerevisae}).   Among these assays,  we consider
12  time-course  experiments profiling  $n=5883$  genes under  various
environmental  changes as  listed in  Table \ref{tab:exp_time_course}.
These expression sets form  12 potential response matrices $\bY$, the column number of which corresponds to the number of time points.
\begin{table}[htbp!]
  \centering
  \begin{tabular}{@{}lc|c@{\hspace{1.9em}}c@{\hspace{1.9em}}c@{}}
    \textit{Experiment}     &     \textit{\#     time    point}     &
    \multicolumn{3}{c}{\it \# motifs selected} \\
    \cline{3-5}
    & & $k=7$ & $k=8$ & $k=9$ \\
    \hline
    Heat shock & 8 & 30 & 68 & 43 \\
    Shift from 37\degres to 25 \degres C & 5 & 3 & 11 & 33 \\
    Mild Heat shock & 4 & 24 & 13 & 23 \\
    Response to $\text{H}_2 \text{O}_2$ & 10 & 15 & 10 & 21 \\
    Menadione Exposure & 9 & 16 & 1 & 7 \\
    DDT exposure 1 & 8 & 15 & 10 & 30 \\
    DDT exposure 2 & 7 & 11 & 33 & 21 \\
    Diamide treatment & 8 & 45 & 25 & 35 \\
    Hyperosmotic shock & 7 & 36 & 24 & 15 \\
    Hypo-osmotic shock & 5 & 20 & 8 & 29 \\
    Amino-acid starvation & 5 & 47 & 30 & 39 \\
    Diauxic Shift & 7 & 16 & 14 & 20 \\
    \hline
    \multicolumn{2}{l}{\it total number of unique motifs inferred} & 87 & 82 & 72 \\
  \end{tabular}
  \caption{Time-course data from \cite{gasch2000genomic} considered for regulatory motif discovery}
  \label{tab:exp_time_course}
\end{table}
Concerning  the predictors,  we consider  the set  of all  motifs with
length  $k$  formed with  the  four  nucleotides,  that is  $\clM_k  =
\set{A,C,G,T}^k$.  There are $p =  |\clM_k| = 4^k$ such motifs. Unless
otherwise stated, the motifs in $\clM$ are lined up in lexicographical
order e.g.,  when $k=2$,  $AA,AC,AG,AT,CA,CC,\dots$ and so  on.  Then,
the $n\times p$ matrix of  predictors $\bX$ is filled such that $X_{ij}$
equals the occurrence  count of motif $j$ in  the regulatory region of
gene $i$.

\subsubsection*{Structure  specification} As  we expect  influential motifs
for a  given condition  to be degenerate  versions of each  other, we  first measure  the similarity  between  any two
motifs from $\clM_k$ with the Hamming defined as
\begin{equation*}
 \label{eq:hamming_def}
 \forall a,b\in\clM_k, \quad \mathrm{dist}(a,b) = \text{card}\{i:a_i\neq b_i\}.
\end{equation*}
For a fixed value of interest $0\leq\ell\leq k$, we further define the \textit{$\ell$-distance matrix}
$\bD^{k,\ell}=(d^{k,\ell}_{ab})_{a,b\in\clM_k}$ as
\begin{equation*}
 \label{eq:distance_matrix}
 \forall a,b\in\clM_k, \quad d^{k,\ell}_{ab} = \left\{
 \begin{array}{ll}
 1 & \text{if $\mathrm{dist}(a,b) \leq \ell$} \\
 0 & \text{otherwise.} \\
 \end{array}\right.
\end{equation*}


$\bD^{k,\ell}$ can be viewed as the  adjacency matrix of a graph where
the nodes are the motifs and where  an edge is present between 2 nodes
when the 2 motifs  are at a Hamming distance less  or equal to $\ell$.
We finally  use the Laplacian  of this  graph as a  structuring matrix
$\bL^{k,\ell}=(\ell^{k,\ell}_{ab})_{a,b\in\clM_k}$, that is
\begin{equation}
 \label{eq:laplacian}
 \ell^{k,\ell}_{ab} =
 \begin{cases}
 \sum_{c\in\clM_k} d^{k,\ell}_{ac} & \text{if } a = b, \\
 -1 & \text{if } d^{k,\ell}_{ab} = 1, \\
 0 & \text{otherwise}.
 \end{cases}
\end{equation}
{Note that we came across a similar proposal by \cite{li2010learning}
  in the context of sparse, structured PLS. The derivation, however, is
  different, and the objective of this method is not the
  selection of motifs but of families of motifs via the compression
  performed by the PLS.  }

\subsubsection*{Results}~  We apply  our methodology  for candidate  motifs
from  $\clM_7,\clM_8$ and  $\clM_9$, which  results in  three lists  of
putative motifs having a direct effect on gene expression.  Due to the
very large  number of potential  predictors that comes with  a sparser
matrix $\bX$  when $k$  increases, we first  perform a  screening step
that keeps  the $5,000$ motifs  with the highest  marginal correlations
with  $\bY$.   Second,  SPRING  is  applied to  each  of  the  twelve
time-course experiments described  in Table \ref{tab:exp_time_course}.
The selection of $(\lambda_1,\lambda_2)$ is  performed on a grid using
the BIC \eqref{eq:pen_crit}.  At the end of the
day,   the  three   lists  corresponding   to  $k=7,8,9$   include
respectively 87, 82 and 72 motifs,  for which at least one coefficient
in the associated  row $\widehat{\bOmega}_{\bx\by}(j,\cdot)$ was found
non-null  for some  of the  twelve experiments,  as detailed  in Table
\ref{tab:exp_time_course}.

To assess the relevance of  the selected motifs, we compared them with
the {\tt MotifDB} patterns available in Bioconductor (\cite{motifdb}),
where known transcription factor binding sites are recorded. There are
453 such reference motifs with size varying from 5 to 23 nucleotides.
Consider the case  of $k=7$ for instance: among  the 87 SPRING motifs,
62 match one  \texttt{MotifDB} pattern each and 25  are clustered into
11 \texttt{MotifDB} patterns as depicted in Table \ref{Tab:MotifDB}.
\begin{table*}[htbp!]
  \begin{center}
    \begin{tabular}{@{}cccc@{}}
      \begin{tabular}{@{}l@{}} 
{\tt CTAAGCCAC} \\ 
\hline 
{\tt ~~TAGCCCC} \\ 
{\tt ~~GCGCCCC} \\ 
\end{tabular} & 
\begin{tabular}{@{}l@{}} 
{\tt GCATGTGAA} \\ 
\hline 
{\tt CCATATG} \\ 
{\tt ~~TTGTGAG} \\ 
\end{tabular} & 
\begin{tabular}{@{}l@{}} 
{\tt CATGTAATT} \\ 
\hline 
{\tt ~~TGTAAAT} \\ 
{\tt ~~TGTATAT} \\ 
\end{tabular} & 
\begin{tabular}{@{}l@{}} 
{\tt TGAAACA} \\ 
\hline 
{\tt TTAGACC} \\ 
{\tt TAAAAAG} \\ 
\end{tabular} \\ 
  \\ 
\begin{tabular}{@{}l@{}} 
{\tt TGATCGGCGCCGCACGACGA} \\ 
\hline 
{\tt ~~~~~~~~~~~GTATAAC} \\ 
{\tt ~~~~~~GCGCCGT} \\ 
\end{tabular} & 
\begin{tabular}{@{}l@{}} 
{\tt TGCTGGTT} \\ 
\hline 
{\tt ~GCTGGTT} \\ 
{\tt ~GCTGGTG} \\ 
\end{tabular} & 
\begin{tabular}{@{}l@{}} 
{\tt GATCGTATGATA} \\ 
\hline 
{\tt ~ATCATAT} \\ 
{\tt ~TTGGTAT} \\ 
\end{tabular} & 
\begin{tabular}{@{}l@{}} 
{\tt ACGCGAAAA} \\ 
\hline 
{\tt ~AACGAAA} \\ 
{\tt ~~ACGAAAA} \\ 
\end{tabular} \\ 
  \\ 
\begin{tabular}{@{}l@{}} 
{\tt CCATACATCAC} \\ 
\hline 
{\tt ~CATAGAC} \\ 
{\tt ~~~~ATATCAC} \\ 
\end{tabular} & 
\begin{tabular}{@{}l@{}} 
{\tt ATTGACCTGGTC} \\ 
\hline 
{\tt ~TCGACTT} \\ 
{\tt ~~CGACTTG} \\ 
{\tt ~~~~~CCAGCTT} \\ 
\end{tabular} & 
\begin{tabular}{@{}l@{}} 
{\tt GACTAGATATATATATTCGAT} \\ 
\hline 
{\tt ~~~~~~~~~~ATATATT} \\ 
{\tt ~~~~~~~CATATAT} \\ 
{\tt ~~~~~~~~~~ATATATG} \\ 
{\tt ~~~~~~~~ATATATA} \\ 
\end{tabular} & 

    \end{tabular}
    \caption{Comparison    of     SPRING-selected    motifs    with
      \texttt{MotifDB}   patterns.   Each   cell  corresponds   to   a
      \texttt{MotifDB}  pattern (top)  compared  to a  set of  aligned
       SPRING motifs with size 7 (down). \label{Tab:MotifDB}}
  \end{center}
\end{table*}
As  seen in  this table,  the clusters  of motifs  selected by  SPRING
correspond  to sets  of variants  of  the same  pattern. These  clusters
consist of motifs  that are close according to  the similarity encoded
in  the structure  matrix $\bL$.   In this  example, the ability of SPRING to
use domain-specific  definitions  of  the
structure between the predictors oriented the regression problem to  account for
motif degeneracy  and helped in selecting motifs that  are consistent
known  binding sites.

\appendices
\section{Proofs}

Most of the  proofs rely on basic algebra and  properties of the trace
and the $\vec$ operators (see e.g. \cite{book_harville}), in
particular
\begin{gather*}
  \trace(\bA^T  \bB  \bC  \bD^T)   =  \vec(\bA)^T  (\bD  \otimes  \bB)
  \vec(\bC),\\
  \vec(\bA\bB\bC) = (\bC^T \otimes \bA) \vec(\bB), \\
  (\bA \otimes\bB) (\bC\otimes\bD) = (\bA\bC) \otimes (\bC\bD).
\end{gather*}

\subsection{Derivation of Proposition \ref{prop:vec_obj}}
\label{sec:vec_obj}

Concerning  the two regularization  terms in  \eqref{eq:objective}, we
have $\|\bOmega_{\bx\by}\|_1 =  \|\bomega\|_1$ since the $\ell_1$-norm
applies element-wise here and
\begin{equation*}
  \trace(\bOmega_{\by\bx} \bL \bOmega_{\bx\by}\bOmega_{\by\by}^{-1}) =
  \bomega^T (\bOmega_{\by\by}^{-1} \otimes \bL) \bomega.
\end{equation*}

As  for the log-likelihood  \eqref{eq:log_lik}, we  work on  the trace
term:
\begin{multline*}
  \trace\left(  \left(\bY +  \bX \bOmega_{\bx\by}\bOmega_{\by\by}^{-1}
    \right)^T \left(\bY          +          \bX
      \bOmega_{\bx\by}\bOmega_{\by\by}^{-1}
    \right)\bOmega_{\by\by}\right)
            = \\ \vec\left(\bY + \bX
      \bOmega_{\bx\by}\bOmega_{\by\by}^{-1}\right)^T
    \left(\bOmega_{\by\by}\otimes\bI_n\right) \vec\left(\bY + \bX
      \bOmega_{\bx\by}\bOmega_{\by\by}^{-1}\right) \\
      =    \left\|\left(\bOmega_{\by\by}\otimes\bI_n\right)^{1/2}
      \left(  \vec(\bY)   +  \left(\bOmega_{\by\by}^{-1}  \otimes  \bX
        \right)
        \vec(\bOmega_{\bx\by}) \right)\right\|_2^2    \\
           =       \left\|     \vec(\bY\bOmega_{\by\by}^{1/2})    +
        \left(\bOmega_{\by\by}^{-1/2}\otimes   \bX   \right)   \bomega
      \right\|_2^2.
\end{multline*}
The rest of the proof is straightforward.

\subsection{Convexity lemma}
\begin{lemma} \label{lemma:yuan_lemma}
The function
\begin{equation*}
-\frac{1}{n}\log L(\bOmega_{\bx\by},\bOmega_{\by\by})      +      \frac{\lambda_2}{2}
  \trace\left(\bOmega_{\by\bx}\mathbf{L}\bOmega_{\bx\by}\bOmega_{\by\by}^{-1}\right)
\end{equation*}
is jointly convex in $(\bOmega_{\bx\by},\bOmega_{\by\by})$ and admits at least one global
  minimum  which  is  unique  when  $n\geq  q$  and  $(\lambda_2\bL  +
  \bS_{\bx\bx})$ is positive definite.
\end{lemma}
The             convexity             of             $-\frac{1}{n}\log
L(\bOmega_{\bx\by},\bOmega_{\by\by})$        is        proved       in
\cite{2014_ITIEEE_yuan}  (Proposition 1).   Similar  arguments can  be
straightforwardly  applied in  the  case at  hand.   Existence of  the
global   minimum   is   related    to   strict   convexity   in   both
$\bOmega_{\bx\by}$     and     $\bOmega_{\by\by}$,    where     direct
differentiation leads to the corresponding conditions.

\subsection{Proof of Theorem \ref{thrm:optim}}
\label{sec:proof_theorem}
The  convexity of criterion  $J(\bOmega_{\bx\by},\bOmega_{\by\by})$ in
$(\bOmega_{\bx\by},\bOmega_{\by\by})$  is  straightforward  thanks  to
Lemma   \ref{lemma:yuan_lemma}   and   considering   the   fact   that
$\|\bOmega_{\bx\by}\|_1$  is  also convex.   One  can  then apply  the
results  developed   in  \cite{2009_MP_tseng,2001_JOTA_tseng}  on  the
convergence  of  block  coordinate  descent for  the  minimization  of
nonsmooth separable  function.  Since \eqref{eq:objective}  is clearly
separable in $(\bOmega_{\bx\by},  \bOmega_{\by\by})$ for the nonsmooth
part  induced   by  the  $\ell_1$-norm,  the   alternating  scheme  is
guaranteed
to converge  to the unique  global minimum under the  assumption of
Lemma \ref{lemma:yuan_lemma}.  It remains to show that  the two convex
optimization subproblems \eqref{eq:optim_cov} and \eqref{eq:optim_par}
can be (efficiently) solved in practice.
\\

Firstly,  \eqref{eq:optim_par}  can  be   recast  as  an
Elastic-Net problem, which  in turn can be recast  as a LASSO problem
(see, e.g. \cite{2005_JRSS_Zou,2010_AOAS_slawski}).  This is
straightforward   thanks  to   Proposition   \ref{prop:vec_obj}:  when
$\hat{\bOmega}_{\by\by}$  is fixed,  solution  to \eqref{eq:optim_par}
can be obtained via
\begin{multline}
  \label{eq:enet_recast}
  \hat{\bomega} \left(= \vec(\hat{\bOmega}_{\bx\by}) \right) \\
  =\argmin_{\bomega\in\Rset^{pq}} \frac{1}{2} \|\mathbf{A}\bomega - \mathbf{b}\|_2^2 + \frac{\lambda_2}{2}
  \bomega^T   \tilde{\bL}
  \bomega + \lambda_1  \|\bomega\|_1 ,
\end{multline}
where $\bA,\mathbf{b}$ and $\tilde{\bL}$ are defined by
\begin{multline*}
\mathbf{A} =
  \left(\hat{\bOmega}_{\by\by}^{-1/2}   \otimes   \bX/\sqrt{n}  \right),
  \\ \mathbf{b} = - \vec\left(\bY\hat{\bOmega}_{\by\by}^{1/2}\right)/ \sqrt{n}
  \text{ and }
  \tilde{\bL} = \hat{\bOmega}_{\by\by}^{-1} \otimes \bL.
\end{multline*}

Secondly, we  can solve  analytically \eqref{eq:optim_cov}
with  simple matrix algebra. By differentiation  of the objective \eqref{eq:objective}  over
$\bOmega_{\by\by}^{-1}$ we obtain the quadratic form
\begin{equation*}
  \bOmega_{\by\by}       \bS_{\by\by}      \bOmega_{\by\by}      -
  \bOmega_{\by\by}                                                =
  \bOmega_{\by\bx}(\lambda_2 \bL + \bS_{\bx\bx})\bOmega_{\bx\by}.
\end{equation*}
After  right  multiplying both  sides  by  $\bS_{\by\by}$, it  becomes
obvious       that      $\bOmega_{\by\by}       \bS_{\by\by}$      and
$\bOmega_{\by\bx}(\lambda_2                    \bL                   +
\bS_{\bx\bx})\bOmega_{\bx\by}\bS_{\by\by}$ commute  and thus share the
same  eigenvectors  $\bU$.    Besides,  it  induces  the  relationship
$\eta_j^2  -\eta_j  = \zeta_j$  between  their respective  eigenvalues
$\eta_j$ and $\zeta_j$,  and we are looking for  the positive solution
of  $\eta_j$.   To  do  so,  first  note that  we  may  assume  that
$\bOmega_{\by\bx} (\lambda_2 \bL + \bS_{\bx\bx}) \bOmega_{\bx\by}$ and
$\bS_{\by\by}$  are  positive  definite, when  $\bOmega_{\bx\by}  \neq
\bzr$  and $n\geq q$;  and second,  recall that if a matrix is  the  product  
of  two positive  definite  matrices then its eigenvalues  are
positive. Hence, $\zeta_j>0$ and the positive solution of $\eta_j$ is $\eta_j = (1 +
\sqrt{1+4\zeta_j})/2$. We thus obtain
\begin{equation}
  \label{eq:cov_prf}
  \hat{\bOmega}_{\by\by} = \bU \diag(\boldsymbol\eta) \bU^{-1} \bS_{\by\by}^{-1}.
\end{equation}
Direct  inversion  yields
\begin{equation*}
  \hat{\bOmega}_{\by\by} = \bU \diag(\boldsymbol\eta/\boldsymbol\zeta) \bU^{-1}
  \hat{\bOmega}_{\by\bx} (\lambda_2\bL + \bS_{\bx\bx}) \hat{\bOmega}_{\bx\by} (=\hat{\bR}^{-1}),
\end{equation*}
To get an expression for $\hat{\bOmega}_{\by\by}$  which  does  not require  additional  matrix
inversion, just note that
\begin{multline*}
  \bS_{\by\by}^{-1} =
  \left(\bOmega_{\by\bx}\hat{\bSigma}_{\bx\bx}^{\lambda_2}\bOmega_{\bx\by}\bS_{\by\by}\right)^{-1}
  \bOmega_{\by\bx}\hat{\bSigma}_{\bx\bx}^{\lambda_2}\bOmega_{\bx\by} \\
   = \bU \diag(\boldsymbol\zeta^{-1})\bU^{-1}\bOmega_{\by\bx}\hat{\bSigma}_{\bx\bx}^{\lambda_2}\bOmega_{\bx\by}
\end{multline*}
where $\hat{\bSigma}_{\bx\bx}^{\lambda_2} = (\lambda_2\bL + \bS_{\bx\bx})$.
Combined with\eqref{eq:cov_prf}, this last equality leads to
\begin{equation*}
    \hat{\bOmega}_{\by\by}^{-1}           =          \bS_{\by\by}\bU
    \diag(\boldsymbol\eta^{-1}) \bU^{-1} (=\hat{\bR}).
\end{equation*}
Finally, in the particular case where $\hat{\bOmega}_{\bx\by}    =   \bzr$,    $\hat{\bOmega}_{\by\by}   =
  \bS_{\by\by}^{-1}$.

\subsection{Derivation of Proposition \ref{prop:df_spring}}
\label{sec:df_spring}

To  apply the  results developed  in  \cite{2012_EJS_Tibshirani} that
rely  on  the   well-known  Stein's  Lemma  (\cite{1981_AS_Stein}),  we
basically need to recast our problem as a classical LASSO applied on a
Gaussian linear regression  model such that the response  vector follows a
normal distribution of  the form $\mathcal{N}(\bmu,\sigma \bI)$.  This
is straightforward by means  of Proposition \ref{prop:vec_obj}: in the
same way as  we derive expression \eqref{eq:enet_recast}, we  can go a
little further and reach the following LASSO formulation
\begin{equation}
  \label{eq:lasso_recast}
  \argmin_{\bomega\in\Rset^{pq}} \frac{1}{2} \left\|
    \begin{pmatrix}
      \mathbf{A} \\ \sqrt{\lambda_2} \tilde{\bL}^{1/2}
    \end{pmatrix} \bomega -
    \begin{pmatrix}
      \mathbf{b} \\ \bzr
    \end{pmatrix} \right\|_2^2  + \lambda_1  \|\bomega\|_1,
\end{equation}
with $\bA,\bb$ and $\tilde{\bL}$ defined as in \eqref{eq:enet_recast}.
From model \eqref{eq:cond_model_mat}, it  is not difficult to see that
$\mathbf{b}$ corresponds to an uncorrelated vector form of $\bY$ so as
\begin{equation*}
  \begin{pmatrix}
    \mathbf{b} \\ \bzr
  \end{pmatrix} =
  \begin{pmatrix}
    -\vec(\bY \bOmega_{\by\by}^{1/2}) \\ \bzr
  \end{pmatrix} \sim \mathcal{N}(\bmu, \bI_{nq}).
\end{equation*}
The  explicit form  of  $\bmu$ is of no  interest here.   The point  is
essentially to  underline that the response vector  is uncorrelated in
\eqref{eq:lasso_recast},  which  allows  us  to  apply  Theorem  2  of
\cite{2012_EJS_Tibshirani} and results  therein, notably for the
Elastic-Net. By these  means, an unbiased estimator of  the degrees of
freedom in \eqref{eq:lasso_recast} can be written as a function of the
active set $\mathcal{A}$ in $\hat{\bomega}^{\lambda_1,\lambda_2}$:
\begin{equation*}
  \hat{\mathrm{df}}_{\lambda_1,\lambda_2} = \trace\left(
    \mathbf{A}_{\mathcal{A}} \left( (\bA^T\bA + \lambda_2 \tilde{\bL}\ )_{\mathcal{A}\mathcal{A}}\right)^{-1}
    \mathbf{A}_{\mathcal{A}}^T
  \right).
\end{equation*}
Routine simplifications lead to the  desired result for the degrees of
freedom.

\section*{Acknowledgment}

We would  like to  thank Mathieu Lajoie  and Laurent  Br\'eh\'elin for
kindly sharing the dataset from \cite{gasch2000genomic}.

\ifCLASSOPTIONcaptionsoff
  \newpage
\fi

\bibliographystyle{IEEEtran}
\bibliography{biblio_spring}

\end{document}